# AI-guided transition path sampling of lipid flip-flop and membrane nanoporation

Matthias Post and Gerhard Hummer

*Max Planck Institute of Biophysics, 60438 Frankfurt am Main, Germany*

(February 17, 2025)

We study lipid translocation ("flip-flop") between the leaflets of a planar lipid bilayer with transition path sampling (TPS). Rare flip-flops compete with biological machineries that actively establish asymmetric lipid compositions. Artificial Intelligence (AI) guided TPS captures flip-flop without biasing the dynamics by initializing molecular dynamics simulations close to the tipping point, i.e., where it is equally likely for a lipid to next go to one or the other leaflet. We train a neural network model on the fly to predict the respective probability, i.e., the "committor" encoding the mechanism of flip-flop. Whereas coarse-grained DMPC lipids "tunnel" through the hydrophobic bilayer, unaided by water, atomistic DMPC lipids instead utilize spontaneously formed water nanopores to traverse to the other side. For longer DSPC lipids, these membrane defects are less stable, with lipid transfer along transient water threads in a locally thinned membrane emerging as a third distinct mechanism. Remarkably, in the high (~660) dimensional feature space of the deep neural networks, the reaction coordinate becomes effectively linear, in line with Cover's theorem and consistent with the idea of dominant reaction tubes.

With advances in experimental techniques, the asymmetry of biological membranes has been receiving increasing attention[1,2]. The plasma membrane, for instance, is highly asymmetric in terms of the lipid composition of its two lipid leaflets[3-5]. While phosphatidylserine (PS) is abundant in the cytosolic inner leaf, its appearance in the outer leaflet indicates a compromised cell membrane, for instance as a result of viral infection, and triggers apoptotic cell destruction[6,7]. To establish this asymmetry against an entropic driving force, elaborate ATP-driven machineries have evolved to actively translocate lipids between the leaflets[8-10] or to trap lipids on one side by covalent modifications such as glycosylation in the Golgi apparatus[11]. Without scramblases[12,13]—a class of proteins that passively redistribute lipids between leaflets—an established leaflet asymmetry tends to persist on biologically relevant timescales. One major reason for this persistence is that spontaneous "flip-flop"[14] of lipids between the two leaflets is rare[15,16]. For flip-flop to occur, the polar or charged lipid headgroup has to pass across the apolar membrane, which is thermodynamically highly unfavorable[17]. A headgroup-dependent enthalpic cost and a tail-length-dependent entropic cost[18] result in small rates of lipid flip-flop that decrease exponentially with bilayer thickness[19].

Lipid flip-flop rates have been measured primarily by labeling lipid headgroups, e.g., with fluorophores and spin-labels[8,19]. Label-free measurements have been limited mostly to challenging neutron-based experiments and sum-frequency vibrational spectroscopy[20]. Despite some differences between label-based and label-free measurements[18], the kinetics of flip-flop is consistently slow. Even for a zwitterionic short-chain lipid such as 1,2-dimyristoyl-sn-glycero-3-phosphocholine (DMPC, C14), spontaneous flip-flops occur only on a minute timescale per lipid[21-23].

Molecular dynamics (MD) simulations promise a label-free view of the lipid flip-flop mechanism[24-26]. In MD, the passage of lipids flipping their membrane orientation can be studied in full microscopic spatio-temporal detail. For instance, previous numerical studies noticed a connection of flip-flop to the formation of transient water pores[24]. Artificially forcing single lipids to move into the bilayer results in water defects, which then may span the whole bilayer[27]. Conversely, creating pores in a membrane (e.g., in case of ionic charge imbalance[28-31], via electroporation[32-34], or lateral/osmotic stress[35]) allows lipids to cross between leaflets by diffusing along the membrane edge around the pore. The free energy cost for pore formation is known to increase with membrane thickness,[36,37] with a trade-off between enthalpy and entropy.[24]

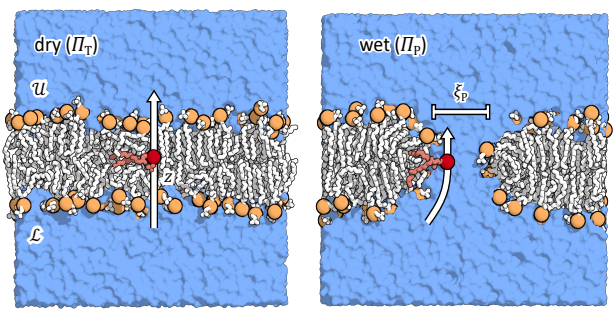

Figure 1. Sketch of the simulation set up and the two initial transition pathways. (Left) The "probe" lipid (red) is pulled from the lower leaflet (state $\mathcal{L}$) to the upper leaflet (state $\mathcal{U}$) to produce a "dry" initial pathway $\Pi_\text{T}$ "tunneling" through the bilayer. (Right) Alternatively, the probe lipid (red) moves along the edge of a pre-established water nanopore in the "wet" pore pathway $\Pi_\text{P}$. Lipids are shown as sticks, phosphorus as orange ball, and water as surface.

Spontaneous lipid flip-flop has thus at least two conceivable, distinct reaction channels – even if one ignores the assistance by membrane protein scramblases or other membrane insertions. In the "tunneling" pathway (denoted $\Pi_\text{T}$ in Figure 1), the phospholipid flips in isolation, with its headgroup passing through the bilayer and its acyl chains reorienting in the membrane. In the pore pathway (denoted $\Pi_\text{P}$), a water-filled pore transiently opens in the membrane and one or several lipids then traverse across the pore-lining membrane edge before the pore closes again. The vertical position $z$ of the headgroup is a natural "reaction coordinate" for transversal displacement. As coordinate for the presence and size of a possibly associated pore, we use $\xi_\text{P}$ by Hub[36], which accounts for the occupancy of polar atoms within the midplane[38,39] and their mean (lateral/axial) distance to the nucleation center.[40]

To resolve the dominant mechanism among multiple pre-identified choices—here direct versus pore-mediated flip-flop—one could try to calculate and compare the respective transition rates. However, this often proves challenging, in particular for a process occurring on the minute timescale. Two common strategies to overcome this difficulty are coarse-grained models[41,42], i.e., representing the lipids and solvent by larger beads, or including a steering bias, e.g., by umbrella sampling.[37] The former coarse-graining is known to result in much faster kinetics and also in less stable water pores due to the simplistic interaction potential and entropy loss[43]. The latter bias may, if poorly chosen, cause inadequate estimates, if degrees of freedom orthogonal to the bias are relevant for the process.

Here, we instead use the recently developed "Artificial Intelligence for Molecular Mechanism Discovery" (AIMMD)[44]. In AIMMD, we apply transition path sampling (TPS)[45] to harvest reactive trajectories without the application of bias forces. TPS is combined with learning the committor[46,47] on-the-fly, encoded in a deep neural network, which results in the identification of important reaction coordinates with close-to-optimal efficiency and with unbiased dynamics.

While very early pioneering work studying lipid flip-flop via TPS only used coarse-grained models,[48,49] AI-guidance in AIMMD allows us to study the molecular mechanism in full detail by sampling hundreds of lipid flip-flop events in atomistic MD simulations. We apply this general framework to DMPC lipid bilayers, as a lipid studied extensively in experiments[22,50-52]. We compare results for atomistic MD simulations with those obtained using Martini coarse-graining. By seeding the AIMMD simulations with initial paths in the two extreme pathways $\Pi_\text{T}$ and $\Pi_\text{P}$, we establish the relaxation of the TPS to the dominant mechanism. In this way, we show that DMPC lipids prefer tunneling in the Martini model and pore-formation in the all-atom MD model. Beyond the mechanism of lipid flip-flop, AIMMD also discovers the mechanism for the spontaneous formation of a membrane pore in a DMPC lipid bilayer, as a combination of a pore-size measure

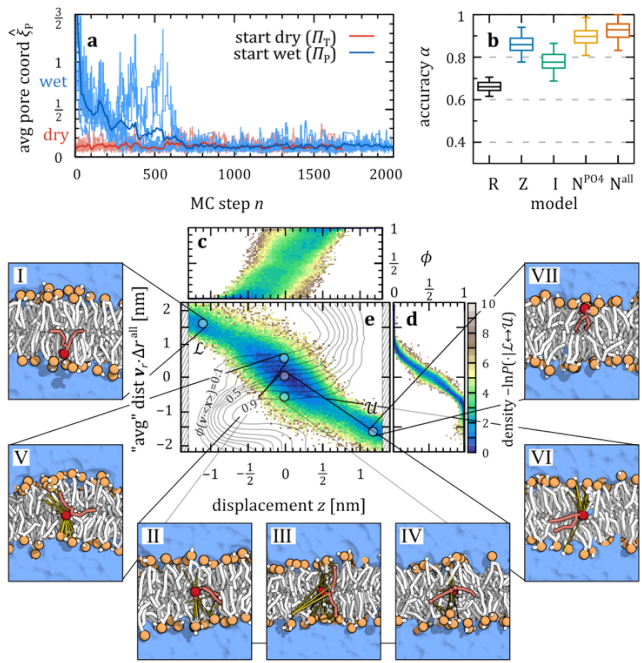

Figure 2. Coarse-grained DMPC lipids tunnel through membrane. (**a**) Time-average of pore reaction coordinate, $\hat{\xi}_P$, evolving during the TPS MC chain. We compare samplers starting from an intact membrane (dry path $\Pi_T$, red) and from a formed pore (wet path $\Pi_P$, blue). Dark colors show a sample average, smoothed over 10 TPS MC steps. (**b**) Accuracy of different committor models, as listed in Table 1. Boxes show median and 50% of bootstrap estimates, whiskers 95%. (**c,d**) Distribution of committor estimates for a given feature, comparing transversal displacement $z$ (c) with a linear combination of bead-to-bead distances to neighboring lipids, $\Delta r^{\text{all}} \cdot v_r$ (e). (**e**) Projection of the TPE onto $z$ and $\Delta r^{\text{all}} \cdot v_r$. Gray iso-lines show the committor with $x$ averaged over 5000 nearest-neighbors (0.6% of all data). Dots are picked as representative configurations corresponding to the snapshots shown in the seven side-panels I–VII. Nearest neighbors are colored in white with increasing intensity, and distances in yellow.

($\xi_P$) and a lipid-displacement ($z_1$). For thicker bilayers formed by long-tailed DSPC lipids, we see how a less stable pore results in a close interconnection of flip-flop and pore formation, where the translocation is catalyzed by an unstable defect, e.g., by moving along narrow water threads across a locally thinned, hour-glass-shaped membrane.

Table 1. Committor network models.

| Model | Description |
|---|---|
| R | Random (uniform) committor assignments |
| Z | Transversal displacement $z$ of lipid probe's head (PO4 bead/P atom) |
| $Z_1$ | Displacement of lipid closest to midplane |
| I | Internal coordinates (see Methods) |
| $N^{PO4}$ | $z$-distance to first 5+5 next-neighboring (NN) lipid's PO4 bead of upper and lower leaflet |
| $N^{all}$ | Distance to all beads of the first 3+3 NN lipids to those of the probe |
| X | Pore reaction coordinate $\xi_P$ (see Methods) |
| D | Depletion of P and N atoms from the membrane measured by the largest distance between 4 NN, $\Delta z^{\text{all}}_{\text{NP1}-4}$ (see Methods) |

## Results

### Martini DMPC lipids prefer tunneling mechanism

We start our investigation with a coarse-grained DMPC lipid model, referring to Methods for details of the TPS setup. The flip-flop transition of an individual lipid (the "probe lipid") is tracked by its transversal displacement $z$ from residing in the lower leaflet (state $\mathcal{L}$) to the upper leaflet (state $\mathcal{U}$). Here, we time-order the TPs as $\mathcal{L} \to \mathcal{U}$, but note that at equilibrium each TP is equally likely to occur in the reverse direction. We measure how the shape of the lipid bilayer changes over the course of the Monte Carlo (MC) chain of transition paths. A TPS MC step here corresponds to one two-way trajectory shooting attempt. Figure 2a shows the (time) averaged pore reaction coordinate, $\hat{\xi}_P$, as a function of the MC step $n$ in TPS, where values $\hat{\xi}_P \gtrsim 1$ indicate the presence of a membrane-spanning water pore. We track each of the samplers starting from $\Pi_T$ (red) and $\Pi_P$ (blue) individually (faint) as well as their mean (solid).

In $\Pi_T$, we initiate the TPS MC chains with an intact flat DMPC double-layer without pore, $\hat{\xi}_P \approx 0.1$. Over the course of the MC chain, $\hat{\xi}_P$ continues to fluctuate around that value. Thus, the transition mechanism remains in $\Pi_T$, i.e., one without the utilization of water pores. By contrast, when starting from $\Pi_P$, the initial artificially large pore rapidly shrinks from $\hat{\xi}_P > 1$ towards $\approx 0.5$. For the first few hundred MC steps, $\hat{\xi}_P$ does not further drop, i.e., the water pore is not fully closed. In this intermediate phase of TPS, the lipid probe is still connected to neighboring water beads, e.g., via a connecting water thread (see Fig. S1a in the SI for an exemplary transition). The connection breaks, though, as the probe reaches the other side, flushing out all water beads of the membrane (see also Fig. S1b). Notably, in this intermediate, unstable period, the corresponding transition times are the smallest, even compared to $\Pi_T$ (see Fig. S1c in the SI). After about $n \approx 500$ MC steps, we have a behavior similar to $\Pi_T$, and thus, all water beads are flushed out and the pore is closed completely during the remaining transitions. There are only rare occasions of single water beads penetrating the membrane, even while the probe is situated in the mid-plane (see also Fig. S1b).

By initializing the MC samplers from the two competing mechanisms $\Pi_T$ and $\Pi_P$ and observing that all TP samplers converged to $\Pi_T$, we clearly see that Martini DMPC lipids prefer to flip-flop without utilizing transient water pores. To further explain how they instead tunnel through the bilayer, we study the importance of individual microscopic features describing the committor $\phi(x)$. To that end, we train on the $\Pi_T$ data, evaluate a variety of neural network models of $\phi$ and measure their respective TP prediction accuracy.

Figure 2b compares how the use of different input features affects the accuracy $\alpha$ of the model; see Methods for its definition. We find that a full description of the tunnel transition mechanism requires the relative distances, $x = \Delta r^{\text{all}}$ (red), of beads of the lipid neighbor network. While the internal state of the probed lipid (green, e.g., splitting its tails) constitutes a poor predictor of TPs, its transversal displacement $z$ (blue) instead does already a reasonable job. Adding direct information about the neighboring lipids, like the vertical position $\Delta z^{\text{PO4}}$ of their PO4 beads, or better the aforementioned $\Delta r^{\text{all}}$, then gives close-to-optimal prediction accuracy. We refer to Methods for details on the network architectures.

When we now train a network model on all these features, and try to understand how they are encoded into $\phi(x)$, we find that the direction $v = \langle \nabla \phi \rangle_\phi / |\langle \nabla \phi \rangle_\phi|$ of the reactive flux averaged on iso-surfaces of $\phi$ hardly changes with increasing $\phi$ (see Fig. S2 in the SI). This implies a simple shape of the committor, $\phi \approx \varphi(x \cdot v)$, i.e., a linear model of the input features. The flux direction $v$ emphasizes the probe's tilt angle[31,53,54] even more than $z$. Yet, most of the weight is found in the distances of the probe head to neighboring lipids (Fig. S2). A possible interpretation of this weighted average of neighbor distances as reaction coordinate may be that the network learned how to better identify the geometry and center of the membrane. But the linearity of this model also implies that there is no particular sequence of events (no specific conformational change of the lipid and its neighbors) resulting in flip-flop.

It is instructive to compare the now identified important mean neighbor distance, $\Delta r^{\text{all}} \cdot v_r$, with the probe's vertical displacement $z$ based on their committor estimates (Figure 2c,d) and the transition path ensemble (TPE) projected onto these features (Figure 2e). The flip-flop starts by inserting the lipid tail-first from one leaflet (state $\mathcal{L}$ in panel I) into the bilayer. It then tilts into the cavity inside the midplane, where a variety of distinct conformations with the same insertion depth $z$ have the same commitment probability (Figure 2c). E.g., a conformation with joined tails (panel II) and with split tails either parallel (panel III) or perpendicular (panel IV) are projected to roughly the same spot, balancing the distances to the two leaflets. Conformations where $\phi$ varies for the same $z$ are then best resolved by the weighted average $\Delta r^{\text{all}} \cdot v_r$ (Figure 2d, and the (gray) iso-lines of $\phi(\langle x \rangle_{z, \Delta r^{\text{all}} \cdot v})$ in Figure 2e). In a snapshot, one then may identify single features differing from the mean to be the deciding factor for commitment, e.g., the probe head starting to penetrate either of the leaflets if single neighboring headgroups are close and accessible (Panel V and panel VI). At the end, the lipid is pushed out straight (state $\mathcal{U}$ in panel VII). See also Fig. S3 for time-traces of exemplary TPs.

## Charmm36 DMPC lipids utilize transient water pores

We repeat the same procedure with an all-atom representation of the DMPC lipids; see Methods. We again classify the overall mechanism of transition by means of the pore defect $\hat{\xi}_P$, as shown in Figure 3a (top). The samplers prepared initially in $\Pi_P$ (blue) all stay in the pore state, $\hat{\xi}_P > 2$. By contrast, the samplers initiated in $\Pi_T$, i.e., with an intact membrane (red), all start with $\hat{\xi}_P$ well below 1. Still, each individual sampler eventually switches to $\hat{\xi}_P \approx 2$ as the TPS MC chain progresses. TPS of the flipping lipid probe thus triggers the formation of water nanopores across the bilayer, which are then kept intact throughout the remaining MC chain. While the probe is able to drag a few water molecules from the get-go (see, e.g., the average neighboring water in Fig. S4a, and Fig. S4b for exemplary trajectories), a switch to a fully connected water chain ($\hat{\xi}_P \approx 1$) is relatively abrupt, meaning the pore is quickly filled during only a few MC steps. The pore then finally relaxes to about twice its initial size, in accordance with Ref. 55 (see also Fig. S5 for exemplary TPs). The open nanopores persist for about 0.4 μs on average (Fig. S6a), far beyond the ≈ 25 ns long flip-flop events (Fig. S6b). AIMMD thus shows that in the atomistic model, DMPC lipid flip-flop is associated with the formation of water nanopores across which the lipids then traverse between the leaflets.

We can quantify the efficiency of the AI-guided path sampling in AIMMD by comparing the number of actually generated reactive trajectories ($n_{gen}$) to those expected for the estimated commitment probabilities ($n_{exp}$, given by the cumulative sum over the estimated TP probability $P(\text{TP}|x) = 2\phi(x)[1 - \phi(x)]$[56]). Figure 3b (solid lines) shows $\eta_{\Delta n} = 1 - \Delta n/n$ as a measure of efficiency, with $\Delta n = n_{exp} - n_{gen}$ and $n$ the number of MC steps. Convergence to $\eta_{\Delta n} \approx 0.88$ indicates a good network model of the committor $\phi$ also for the atomistic MD simulations.

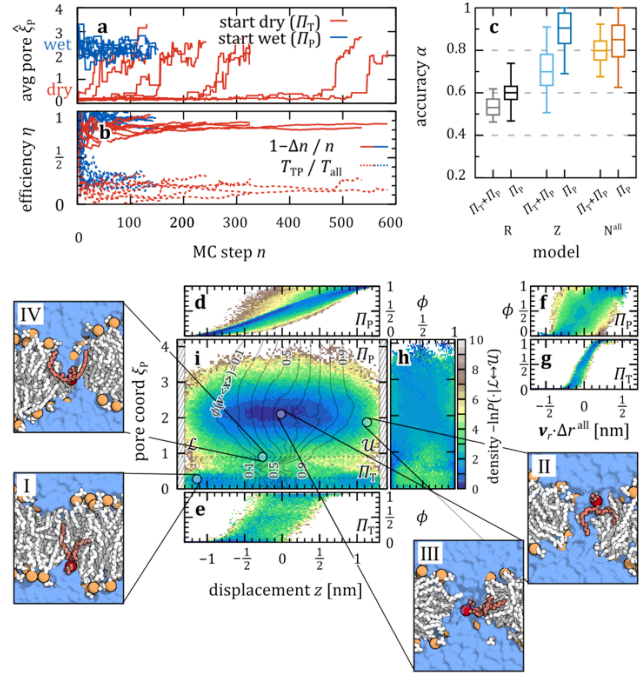

Figure 3. Atomistic DMPC lipids flip bilayers through water-filled nanopores. (**a**) Time-average of $\xi_P$ during the MC chains, comparing sampler starting from the tunnel mechanism, (dry, $\Pi_T$, red), with those starting with a pore (wet, $\Pi_P$, blue). (**b**) Efficiency $\eta$ measured by difference of expected and generated TPs, $\Delta n = n_{exp} - n_{gen}$, and by the simulation time $T_{TP}$ of new transition events compared to the total simulation time $T_{all}$. (**c**) Accuracy of committor models, as listed in Table 1, training on all data (bright), compared to only $\Pi_P$ (dark). Boxes show median and 50% of bootstrap estimates, whiskers 95%. (**d**–**h**) Distribution of committor estimates for a given feature, comparing $z$ (**d**,**e**) and a linear combination of bead-to-bead distances (**f**,**g**), $\Delta r^{all} \cdot v_r$, stratified to the $\Pi_P$ (**d**,**f**) and $\Pi_T$ (**e**,**g**) data, and the pore coordinate $\hat{\xi}_P$ (**h**). (**i**) Projection of the TPE onto $z$ and $\xi_P$. Black iso-lines show the committor with $x$ averaged over 5000 nearest-neighbors (0.6% of all data). Representative configurations are shown in the four side-panels I-IV.

We alternatively measure the efficiency by $\eta_T = T_{TP}/T_{all}$ comparing the aggregate time $T_{TP}$ of newly accepted transition paths entering the TPS Markov chain to the total simulation time $T_{all}$ (Figure 3b, dashed lines). With ~18% of MC steps resulting in accepted TPs (16% of $\Pi_T$, 23% of $\Pi_P$) of a combined time $T_{TP} \approx 12$ μs, we achieve an efficiency of $\eta_T \approx 0.25$, i.e., a quarter of the time goes to simulating new transition paths.

The analyses of the TPS data and the committor model trained on the shooting results now depend on whether or not we include the large portion of initial $\Pi_T$ transitions. If we do, see Figure 3c, we see that again the committor prediction using distances $\Delta r^{\mathrm{all}}$ to all neighboring lipid atoms (yellow) outperforms a simple model using $z$ alone (light blue). With the smaller sample size, the model accuracy is worse than in the coarse-grained case, but we again see the importance of the precise relative position to the probe's neighbors for a successful transition.

If we, however, consider only the data from the dominant path-type $\Pi_P$ to which all TPS walkers relax eventually, the transversal displacement $z$ (dark blue) suffices to describe the transition mechanism. There is no improvement by using more features, like the distances to neighbors, $\Delta r^{\mathrm{all}}$ (red). While we expect that the AIMMD efficiency would slightly improve when continued sampling, our collected data (1419 $\Pi_P$ paths for in total 3500 shooting points) is convincing enough to confirm the diffusion along $z$ via $\Pi_P$. We also refer to Fig. S7a,b for cross-validation of these models.

We again find that the training process resulted in learning a linear combination of the input features and a uni-directional reactive flux (see Fig. S8 in the SI). In Figure 3d-i we break down its main contributors: the displacement $z$ (d,e; with more weight compared to Martini), and the distance average $\Delta r^{\mathrm{all}} \cdot \boldsymbol{v}$ (f,g; with very similar weight); and compare it to the pore-defining reaction coordinate $\hat{\xi}_P$ (h). To no surprise, we see differences in prediction accuracy of these features depending if we limit the analysis to either the $\Pi_T$ or $\Pi_P$ data. That is, in case of $z$, we see a relatively sharp distribution of $\phi$ when traversing along $\Pi_P$ (Figure 3d), again indicating that $z$ is capable to describe the diffusion. For the $\Pi_T$ mechanism, however, $\phi(z)$ broadens (Figure 3g), which means $z$ is a poor descriptor of the pore-less flip-flop, in line with our Martini result. Conversely, if we look at the weighted average of distance between atoms of probe and neighboring lipids, $\Delta r^{\mathrm{all}} \cdot \boldsymbol{v}_r$, we see a broad distribution corresponding now to $\Pi_P$ (Figure 3f), and a sharp distribution in $\Pi_T$ (Figure 3g). This tells us that $\Delta r^{\mathrm{all}} \cdot \boldsymbol{v}_r$ takes a similar role as in the Martini case, describing the pore-less tunneling via $\Pi_T$ and becoming obsolete after TPS has converged to $\Pi_P$, where $z$ alone suffices. Including $\Delta r^{\mathrm{all}}$ again improves the localization of the effective membrane center of an intact membrane, but not when situated in a pore, as there is no drastic change of the probe neighbors. The state $\xi_P$ of the water pore, in contrast, is always a poor predictor and is thus not deemed important by the model. Figure 3e shows the TPS data projected onto the displacement $z$ and pore shape $\xi_P$. The bottom region, $\xi_P < 1$, represents early trajectories traversing from $\mathcal{L}$ to $\mathcal{U}$ via $\Pi_T$ (see also panel I). Conversely, the upper $\xi_P > 1$ region shows the trajectories starting and ending with a formed open pore, with fluctuations around $\xi_P \approx 2$ due to pore expansion and contraction (panel II and III). The transition from $\Pi_T$ to $\Pi_P$ paths in the TPS MC chain itself appears to be a rare event, associated with the nucleation of a water nanopore in a single trajectory, as reflected in a step in $\xi_P$ (panel IV). Here, the probe's head within the membrane attracts the surrounding water to seed and eventually form a percolating water thread. The iso-lines of $\phi$ projected onto $z$ and $\xi_P$ expand from the narrow $\Pi_T$ to a broader and less committed behavior along $\Pi_P$, but stay roughly parallel to $\xi_P$ otherwise (see also Fig. S7c for a study of the midplane-symmetry).

Without imposing a sealed membrane in the initial and final state, the network model thus did not need to learn the actual transition mechanism of flip-flop, but only the intermediate, diffusive step (along $z$), before and after the formation and closing of the water pore ($\xi_P$). This is because we defined the states $\mathcal{U}$ and $\mathcal{L}$ only via the displacement $z$ such that we observe pore nucleation only during the initial equilibration phase of TPS.

**Flip-flops are rare during pore nucleation**

Therefore, we now aim to capture the nanopore nucleation step prior to the lipid traversal for a full description of the flip-flop process. So far, nucleation of water pores was achieved by merely shooting close to the transition state, hinting at the importance of flip-flopping lipids as seeds for the formation of transient water pores. These nucleation events are now used as TPS starting points, and evaluated via the pore reaction coordinate $\xi_P$ to define the flat and porous membrane. We refer to Methods for simulation details.

We see in Figure 4a that AIMMD is capable of efficiently sampling also pore nucleation. With efficiencies of $\eta_{\Delta n} \approx 0.95$ and $\eta_T \approx 0.24$, we have a total of 256 distinct TPs to analyze the mechanism of pore nucleation. Since $\xi_P$ is treating all lipids as a group, instead of having one tagged lipid probe, we look at the behavior of each lipid, sorted, e.g., by their distance $z_i$ from the midplane.

To elucidate the mechanism of pore nucleation, we again compare different features as inputs for the committor network model, Figure 4b. With further details in Methods, we compare using only $\xi_P$ (model X) with using $z_1$ as inputs (model $Z_1$). Inspired by the work of Ref. 40, we also use the largest depletion of 4 P and N atoms from the nucleation center, $\Delta z_{NP1-4}^{all}$ (model D). While there is a hint of better accuracies $\alpha$ using the latter, both $\xi_P$ and $\Delta z_{NP1-4}^{all}$ do a decent job in predicting TPs. There is no major improvement in $\alpha$ by using more input features, which confirms their suitability as reaction coordinates.

To study the connection of pore nucleation to lipid flip-flop, we start by simply counting all observed translocation events. During most nucleation transitions, no lipids flip. We only observe 10 distinct flip flop events, in $\approx 3\%$ of the TPs. See Fig. S9a,b for an example of these transitions. In all of these cases, the flip-flop is preceded with bulging of the membrane and then water forming a percolating thread between the leaflets (see Fig. S9c for evaluation of $z_1$ compared to the pore in these cases). So, while our shooting point selection in the previous sampling of lipid flip-flop inevitably resulted in the nucleation of water-pores, we can rule out lipid flip-flop as a main, native trigger of pore nucleation. Instead, the flip-flop happens at a later stage, with the spontaneously formed nanopores staying open for about 0.4 μs on average, with typically about 15 flip-flop events (Fig. S6a).

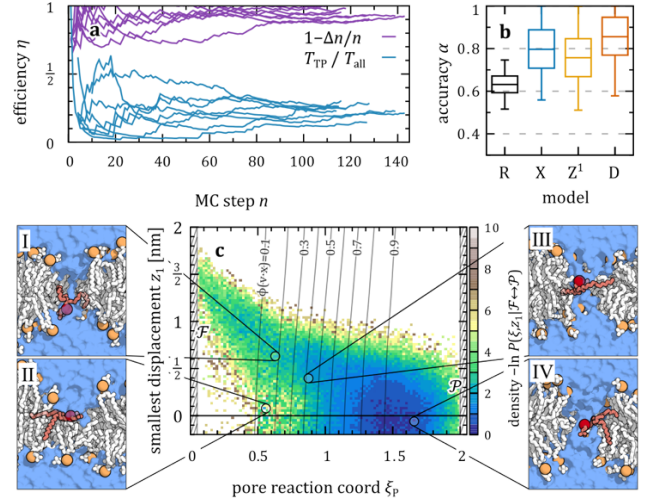

Figure 4. TPS of water nanopore nucleation in membranes of Charmm36 DMPC lipids. (a) Efficiency of AIMMD measured by the difference of expected and generated TPs, $\Delta n = n_{exp} - n_{gen}$, and the ratio between simulation time of new transitions, $T_{TP}$, and total simulation time, $T_{all}$. (b) Accuracy of committor models, as listed in Table 1. Boxes show the median and 50% of bootstrap estimates, whiskers 95%. (c) Pore nucleation mechanism. The center plot shows the TPE for nucleation of a pore (state $\mathcal{P}$) starting from a flat membrane (state $\mathcal{F}$), projected onto the plane spanned by $\xi_P$ and $z_1$. Side panels show representative structures.

By sampling the nucleation process, still, some lipids have to migrate towards the bilayer midplane. The TPE in terms of $\xi_P$ and $z_1$, see Figure 4c, shows how (at least) the lipid closest to the nucleation center migrates into the pore (in comparison to tilted Figure 3g). We see that starting from a flat surface, state $\mathcal{F}$, the membrane starts to bulge and thin locally, thus also changing $z_1$ closer to the midplane. As $\xi_P$ reaches 0.5, a water connection to the other side forms. The insertion of water molecules is then followed by the polar lipid heads, see Fig. S9d, in accordance with Refs. 35,40,57. A pore is formed for $\xi_P > 1$, which then has to stretch to a slightly expanded shape to reach state $\mathcal{P}$.

**Transient water threads and local membrane thinning as third mechanism for flip-flop through thick membranes**

For thick membranes formed by long-tailed DSPC lipids, yet another mechanism emerges: flip-flop mediated by transient and narrow water threads associated with local membrane thinning. We initiate AIMMD simulations of atomistic DSPC lipid bilayers from $\Pi_T$ and $\Pi_P$ initial pathways (Fig. S10 in the SI). In the $\Pi_P$ sampler, the water pores quickly become narrow, and collapse almost immediately after completed flip-flop. This collapse is consistent with water nanopores being disfavored in thick bilayers.[37,55,58] By contrast, the initial "dry" flip-flop in the $\Pi_T$ samplers quickly changes to incorporate water, where the probe head drags individual water molecules as a single shell to the other side. In one case, the flip-flop mechanism transitions to a narrow pore that then persists. Visual inspection shows that the process is initialized by bulging, resulting in local thinning of the membrane, after which a transient water thread[59] forms (see the examples in Fig. S11). The flipping lipid then connects the two DSPC leaflets, which adopt a shape resembling a conic intersection. However, more extensive TPS would be needed for a full quantification of the reactive flux carried by this third reaction channel, intermediate between the "wet" and the "dry" pathways with and without fully formed water nanopores.

## Discussion

Applying the recently developed AIMMD methodology, we conducted an extensive numerical study of unassisted flip-flop of DMPC lipids. We studied the transverse motion of a single lipid through a bilayer of lipids, comparing coarse-grained with all-atom simulation.

We found that the TPE of the atomistic lipids naturally converged to a pore-mediated diffusion mechanism. Once formed, the water nanopores stayed intact during the whole flip-flop process. The emergence of nanopores for all TPS walkers and their persistence provide strong evidence for the dominance of the pore pathway for lipid flip-flop in our atomistic DMPC membrane.

AIMMD was also able to effectively sample water nanopore nucleation in an unbiased way. It confirmed that, first, the pore is established by a percolating water thread[59], which then allowed lipid headgroups to enter into the bilayer, with or without flip-flop. Pore formation thus appears to precede flip-flop, which occurs by chance in pores that live long enough, about 0.4 μs before pore collapse in our system.

We saw the other extreme case of pore-less flip-flop via the $\Pi_T$ pathway with the coarse grained DMPC Martini lipids and at the start of the all-atom TPS MC chains. Despite the involvement of an entire lipid patch in the flip-flop process, the committor network model was able to encode all relevant microscopic details. We found that to best predict the outcome of a ($\Pi_T$) transition, the model needed to take into account the surrounding network of lipids, most importantly their headgroups.

Most strikingly, we found that, after extensive training, our deep neural network with a ~660-dimensional feature space encoded the committor in a nearly linear fashion. While neural networks are in general considered to be quasi black boxes able to approximate highly non-linear and hard-to-interpret functions, our network instead converged to a weighted average of distances to the neighboring lipids as an optimal reaction coordinate, associated with a simple uni-directional reactive flux. This thought-provoking result connects to early linear models for $\phi$[47], as well as the idea of transition tubes[60] as approximately straight pathways through the transition region. The unanticipated tendency to linear models in sufficiently high dimensions is consistent with Cover's theorem[61] as a statement on the effectiveness of linear classifiers in high-dimensional spaces. By increasing the dimension of the feature space, linear models become more effective in discriminating configurations, here according to their committor values. However, the need to regularize the network representation of $\phi$ to prevent overfitting may play a role as well. How linear the transition funnel is close to the transition state emerges as an interesting future research direction well suited for the AIMMD method.

So, which transition mechanism of flip-flop is the correct one? Both atomistic and coarse-grained force fields are known to suffer from inaccuracies, which here may lead to the observed qualitatively different behavior of tunneling, with or without passenger water molecules, and pore mediated flip-flop. For DMPC lipids, the Martini case showed us how a lipid may flip without water but the all-atom representation instead leads to fully grown water pores to diffuse through. The middle-ground might thus be what we observed for atomistic DSPC lipids, where local membrane thinning combined with narrow water threads to establish a passageway for lipid flip-flop.

From here having captured tunnel, pore, and water-thread mechanisms of flip-flop in closely related systems, we deduce that the relevant free energy barriers have comparable heights. The dominance of one or the other mechanism will then depend on system and condition, in line with earlier MD studies (see, e.g., Refs. 24,31,37). For instance, lipids with large polar or highly charged headgroups may favor water nanopores even in a thick bilayer, where tunneling or water threads may dominate for a zwitterionic lipid.

The connection between water pores and flip-flop is also coming into focus in experimental studies (see, e.g., Refs. 16,62,63), having clear ramifications on the mechanistic interpretation of observations of lipid flip-flop-associated relaxation processes[19]. We expect that flip-flop mediated by water nanopores is essentially independent of headgroup size and charge of the flipping lipid. By contrast, flip-flop through dry tunnels should depend strongly on the size and charge of the headgroup, which partially loses its solvation shell during passage through the bilayer. By varying head-groups of the probe lipid and acyl-chain lengths of lipids in the background membrane, it should thus be possible to probe the transitions between different mechanisms of flip-flop, e.g., by estimating the entropy and enthalpy associated with defect density changing with temperature[16].

## Methods

### Molecular dynamics simulation

For MD simulations of the coarse-grained DMPC bilayer, we used gromacs version 2022[64] and the Martini 3[42,65] model (see Fig. S12b,c in the SI for a sketch). A bilayer of 2 × 225 lipids was solvated in water with 0.15 NaCl in a ~12 × 12 × 12 nm³ box using the insane.py[42] script. After energy minimization via gradient descent, the system was shortly equilibrated for 0.1 ns with 2 fs timestep, after which we performed a longer equilibration run with 20 fs time step for 1 μs in the semi-isotropic $NP_{xy}P_zT$ ensemble, using v-rescale thermostat[66] at 310.15 K with $\tau = 1$ ps (membrane and solvent coupled separately), and pressure couplings via Parrinello-Rahman[67] at 1 bar with $\tau = 12$ ps and $\kappa = 3 \times 10^{-3}$ bar$^{-1}$. Van der Waals interactions were handled with cutoff at 1.1 nm with potential-shift, Coulomb interactions via reaction field[68] with r = 1.1 nm with a dielectric constant of 15 and an infinite relative reaction-field dielectric.

We also built a solvated all-atom DMPC bilayer using CHARMM-GUI[69,70], maintaining the initial a 12 × 12 × 12 nm³ box and using TIP3P water with 0.15 NaCl ions (Fig. S12a). The DMPC double layer was modeled by the Charmm36 force field[71]. The CHARMM-GUI schedule was set to a gradient decent minimization with position restraints of the lipids ($k = 1000$ kJ mol$^{-1}$nm$^{-2}$) and their joint dihedral ($k = 1000$ kJ mol$^{-1}$rad$^{-2}$), which was followed by an $NVT$ equilibration with the same restraints for 125 ps with 1 fs time step, with Berendsen thermostat[72] at $T = 310.15$ K with $\tau = 1.0$ ps (membrane and solvent coupled separately) and constrained hydrogen bonds (LINCS[73]). Van der Waals interactions were handled with cutoff at 1.2 nm, with force-switching from 1 nm, Coulomb interactions via PME[74] with $r = 1.2$ nm. Then followed 125 ps with $k = 400$ kJ mol$^{-1}$nm$^{-2}$ and

400 kJ mol$^{-1}$rad$^{-2}$, respectively, after that a 125 ps $NP_{xy}P_zT$ run at 1 bar with $\tau = 5$ ps and $\kappa = 4.5 \times 10^{-5}$ bar$^{-1}$, with $k = 400$ kJ mol$^{-1}$nm$^{-2}$ and 200 kJ mol$^{-1}$rad$^{-2}$, then 125 ps with 2 fs time step and $k = 200$ kJ mol$^{-1}$nm$^{-2}$ and 200 kJmol$^{-1}$rad$^{-2}$, then 125 ps with $k = 40$ kJ mol$^{-1}$nm$^{-2}$ and 100 kJmol$^{-1}$rad$^{-2}$, and then 125 ps without restraints. We then performed a 100 ns long simulation with 2 fs time step, v-rescale temperature coupling and Parrinello-Rahman pressure coupling.

## Transition path sampling

To test whether lipids prefer a spontaneous tunneling through the bilayer ($\Pi_T$), or the diffusion through formed water pores ($\Pi_P$), we set up initial transition pathways for these two cases. Pathway $\Pi_P$ required the preparation of a water pore by introduction of a flat-bottomed position restraint of the lipids in the center of the simulation box. To this end, we performed 1 ns (10 ns in case of Martini) of simulations with $k = 500$ kJ mol$^{-1}$nm$^{-2}$ and $r$ ranging from 0.5 (head) to 1.6 nm (tail) to open th tunneling through the bilayer ($\Pi_T$), or the diffusion through formed water pores ($\Pi_P$), we set up initial transition pathways for these two cases (Figure 1, $\Pi_P$). With fixed pore, we performed 10 ns (100 ns in Martini MD) of simulations, in which we also fixed the probe lipid fixed in the middle of the bilayer using an additional cylindrical harmonic restraint of the PO4 group with $r = 2$ nm, $k = 1000$ kJ mol$^{-1}$nm$^{-2}$. We used the last 1 ns (10 ns for Martini) to pool shooting points for parallel TPS using AIMMD (see below). Using 8 samplers (6 for Martini), we ran a total of 100 MC TPS steps, i.e., 100 TPS simulations with fixed pore but unbiased probe lipid, of which we used for each sampler the last accepted one as seed for the following unbiased TPS. See Fig. S12d for snapshots of one of these initial $\Pi_P$ paths. Preparation of initial $\Pi_T$ trajectories was achieved by a harmonic constraint pulling the probe lipid headgroup with $v = 0.001$ nm/ps, $k = 1000$ kJ mol$^{-1}$ nm$^{-2}$, by simultaneously preventing water to enter the double-layer by use of a flat-bottomed position restraint of $k = 500$ kJ mol$^{-1}$nm$^{-2}$ and $r = 1$ nm from the mid-plane, resulting in a pore-free transition (Figure 1, $\Pi_T$). We repeated this procedure to pull both upwards and downwards to have $4 + 4$ ($3 + 3$ for Martini) seed paths. These rough transition pathways were then used for sequential TPS shooting. We ran a total of $N = 1000$ MC steps with water restraint and unbiased probe lipid. We used the last accepted one as seed for the following unbiased TPS. See Fig. Fig. S12e for snapshots of one of these initial $\Pi_T$ paths. In both cases of initial starting transition pathways, we then performed unbiased (i.e., without flat-bottomed restraints) simulations to sample the transition state ensemble. We performed sequential two-way shooting TPS simulations via the AIMMD framework[44].

AIMMD aims for a high success rate of sampling flip-flop transitions by simultaneously estimating the corresponding committor $\phi(x|w)$ via a neural network with weights $w$. We predict from a set of initial microscopic input features $x$ in what state the trajectory will end and from what state it came by minimizing the negative log-likelihood of shooting outcomes,

$$L(\boldsymbol{w}) = \sum_{i=1}^{N} \ln\left[\binom{n}{k_i} \phi(\boldsymbol{x}_i^{\text{SP}}|\boldsymbol{w})^{k_i} \left(1 - \phi(\boldsymbol{x}_i^{\text{SP}}|\boldsymbol{w})\right)^{n-k_i}\right],$$

in terms of the weights $w$ of the network, where for two-way shooting we have $n = 2$ and $k_i \in \{0,1,2\}$, as the number of times a trajectory hits the final state. To accelerate the learning of the committor, we include the SPs of the initial restraint runs to the training set. We produce $N = 1000$ MC steps for $\Pi_P$ (12000 MC steps in case of Martini) and 2500 for $\Pi_T$ (10000 Martini). The estimate of $\phi$ is then used in the sequential TPS to efficiently sample SPs from the previous transition path. We allow for some deviations of shooting from the optimal $\phi = 0.5$ iso-surface by sampling from a Cauchy distribution of the logit $q$ of $\phi$ ($q = \ln\left[\frac{\phi}{1-\phi}\right] \sim$ Cauchy($\mu=0$, $\gamma=1$)). To this end, we estimate the actual distribution $P(q|\text{TP})$ of the TPS data (by a histogram of $q$), to reweigh each frame to a Cauchy sample. We choose SPs uniformly first, create the histogram after 100 MC steps, and update every 250 steps.

The TPS of pore nucleation were seeded by extracting TPs from the $\Pi_T$ samplers transitioning to the $\Pi_P$ mechanism. We first sampled 1000 snapshots uniformly from all trajectory frames with pore reaction coordinate $0.5 < \xi_P < 1.0$. We then used AIMMD to uniformly pick one of these frames until a first trajectory was accepted. After that, we again performed sequential shooting, using 8 samplers with a total of 1000 MC steps, with the same shooting point selection criterion as before.

**Input features and network architecture**

Using MDAnalysis[75], we define the two final states of the transition by the transversal displacement $z$ from the midplane (defined by the lipid P atoms (PO4 bead for Martini) w.r.t. the vertical center ($z = 0$) of all P's), namely the state $\mathcal{L}$ when $z < -1.3$ nm, and state $\mathcal{U}$ when $z > 1.3$ nm, respectively.

To monitor the internal conformations, in addition to $z$, we also tracked the probe lipids radius of gyration and its tilt angle $\theta$ defined by the average distance vector to the P (PO4) atom and the $z$-axis. For interactions with the other lipids, we also recorded the indentation of the upper and lower leaflet by the respective mean squared-displacement from their center.

We then tracked the displacement of each P (PO4) bead w.r.t. the probe sorted by distance, tracking the first 20 neighbors ($3 \times 20$ coordinates) to reduce noise. We also included the number of water molecules in the first and second shell around the probe P (PO4 in Martini), using the indicator function of Ref. 76. As for the total number of water molecules inside the bilayer, we counted the number of water oxygens (W beads) within $\Delta z = 0.5$ of the mid-plane. See a detailed list of input features in Table S1.

During the AIMMD runs, the neural network estimates the committor to the $\mathcal{U}$ state via a latent space representation $\phi(x)$, where we first selected 68 input features $x$ to be encoded through 5 hidden layers. More specifically, the data is processed via a linear compression (with a small dropout probability during training), after which followed a ResNet[77] unit (with ELU activation) of depth 4. We do this to sequentially go from $68 \rightarrow 46 \rightarrow 31 \rightarrow 21 \rightarrow 14 \rightarrow 10 \rightarrow 1$, where at the last step we only use a linear unit. The output $q(x)$ is then transformed by a softmax to the probability $\phi$. See Fig. S13a for a sketch and Table S2. Note that the networks with ~660 features discussed in Results were used later in postprocessing, as described below.

The pore nucleation transitions were defined by the pore reaction coordinate $\xi_P$ from Ref. 55, which combines the process of pore nucleation with that of pore expansion. The former is evaluated in terms of what fraction of the membrane (in terms of slabs along $z$ at the nucleus) is already occupied by polar atoms, the "pore-chain" $\xi_{ch}$[39]. The latter counts the number of water molecules inside a formed (assumed cylindrical) pore to estimate its radius $R$, and is added to $\xi_{ch}^s$ when close to 1, in units of the radius $R_0$ of a just fully nucleated pore. Here, we set the state boundary of the flat membrane, $\mathcal{F}$, to $\xi_P < 0.05$, and that of an expanded pore, $\mathcal{P}$, to $\xi_P > 2.0$. To study which features best describe the committor, we chose as input features of its neural network model all of $\xi_P$ and its constituents $\xi_{ch}$ and $R$.

The reported parameters for DMPC lipids induced an artificial meta-stable state in the transition region; see Fig. S14 and its caption. In case of Martini, we thus changed the parameters of $\xi_P$ by decreasing the number of subdivisions to 4, with a cylinder size of $Z_{\text{mem}} = 1.8$ nm, $R_{\text{cyl}} = 1.0$ nm, counting the polar atoms for calculating the pore radius within $D = 1.2$ nm, as well as changing the switch towards pore expansion at $\xi_{ch}^s = 0.9$ with a radius $R_0 = 0.38$ nm. In this way, we aim to balance the noise around $\xi_P \approx 0$ with being able to detect water threads, as well as a smooth transition for large $\xi_P$.

We also feed in coordinates suggested by Bubnis and Grubmüller[40], who consider the distances of different atom types to the pore center. In our case, we use the pore center definition of Ref. 39, a weighted circular mean of the headgroups. For the isotropic, lateral and axial distance to the center we measured the 1st, 2nd and 3rd NN, as well as an average over the first 2, 3, 4, 5 and 10. For the axial distance, as detailed in Ref. 40, we took the maximal average over neighboring pairs. We chose the same atom types, water O, P, N+P, N+P+$O_{H2O}$, carbon tails, and all carbons. In total, we end up with a network shape $147 \to 85 \to 50 \to 29 \to 14 \to 17 \to 1$. See also Table S3.

**Accuracy of committor models**

After the TPS production run, we tested if network architectures other than the initial one used in AIMMD resulted in a better committor estimate. To this end, we define the accuracy $\alpha$ of the committor model by the excess variance of committor estimates not explained by a binomial distribution. The probability $p_{\text{bin}}$ of $k$ hits of the final state with $n$ shots from a starting configuration with exact committor $P$ is

$$p_{\text{bin}}(k|n, P) = \binom{n}{k} P^k (1-P)^{n-k}.$$

We assume that $P$ is beta distributed around our estimate $\phi$ of the committor,

$$p_{\text{beta}}(P|a,b) = \frac{1}{B(a,b)} P^{a-1}(1-P)^{b-1},$$

which defines the Bayesian conjugate prior normalized by the beta function $B(a,b)$. We enforce the means to match, $\langle P \rangle = \phi$, by setting

$$a = \frac{\alpha}{1-\alpha}\phi, \qquad b = \frac{\alpha}{1-\alpha}(1-\phi),$$

with a constant $\alpha$ in the range $0 \leq \alpha \leq 1$. The variance of $P$ in the beta distribution is then

$$\text{Var}[P] = (1-\alpha)\phi(1-\phi),$$

with its maximum and minimum at $\alpha = 0$ and 1, respectively. Convolving the binomial and beta distributions gives the probability to see $k$ hits,

$$p(k|n, \phi, \alpha) = \int_0^1 dP\, p_{\text{bin}}(k|n,P) p_{\text{beta}}(P|a,b)$$
$$= \binom{n}{k} \frac{B(w\phi + k, w(1-\phi) + n - k)}{B(w\phi, w(1-\phi))},$$

which is a beta-binomial distribution of $k$, where $w = \alpha/(1-\alpha)$.

We now treat $\tilde{p}(\alpha|k,n,\phi) \propto p(k|n,\phi,\alpha)$ as a Bayes posterior for the accuracy $\alpha$, having treated $P$ as a nuisance parameter.

Given a sample of shooting data, $(\phi_i, k_i, n_i)_{i=1}^N$, where $\phi_i$ is the committor predicted by the model, we accordingly estimate the accuracy $\alpha$ of the committor model by maximizing the log-posterior,

$$L(\alpha) = \sum_{i=1}^N \ln[p(k_i|n_i, \phi_i, \alpha)],$$

with respect to $\alpha$. For $\alpha = 1$, the committor model fully explains the data, $\phi_i = P_i$ for all $i$; for $\alpha = 0$, the data are best explained by a combination of fully committed states, $P = 0$ and $P = 1$, indicating a complete lack of predictive power.

We make a bootstrap estimate of $\alpha$ by repeating 10 times: split the data into training (all but one MC chain) and validation set (that chain), train the model for some number of epochs, and then pick 100 times bootstrap samples (with replacement) from the validation set to estimate the validation loss and accuracy. We used the validation loss to set the number of epochs to prevent over-fitting. We tested the influence of the number of hidden layers, number of nodes, as well as dropout. See Table S2 and Fig. S13b for all tested networks.

As a final test of systematic error of the network's predictions $\phi_i$, we performed additional committor shots now with $n_i = 20$ (instead of the $n_i = 2$ during TPS) for an estimate of the actual committor, $P \approx k_i/n_i$ (Fig. S7a,b). To assess bias and variance, we try to avoid the binning used previously[44], and instead fit a line to the actual logit ($\ln k/(n-k)$) vs. the logit predicted by the model ($\ln \phi/(1-\phi)$),

$$q^{\text{lin}} = \ln\frac{\phi^{\text{lin}}}{1-\phi^{\text{lin}}} = c\ln\frac{\phi}{1-\phi} + d, \quad c_{\text{fit}}, d_{\text{fit}} =$$
$$\arg\min_{c,d} \sum_{i=1}^N \left(q_i^{\text{lin}} - \ln\frac{k_i}{n_i-k_i}\right)^2.$$ Together with the estimate for $\alpha$, we can visualize the spread of the data by the standard error of the mean,

$$\Delta \phi^{\text{lin}} = \sqrt{1-\alpha\left(1-\frac{1}{n}\right)} \sqrt{\phi^{\text{lin}}(1-\phi^{\text{lin}})}.$$

We also tested additional input features of the network not used in the initial network models

(which had used only 68 features). While in the initial model we simply averaged all P (PO4) atoms to define the midplane, we got better predictions by using the lipid tail atoms weighted (using the sigmoid of Ref. 78) by their distance to the probe in the $xy$-plane, see Fig. S13b. We also refined the definition of the tilt angle $\theta$ by calculating the normal to the membrane via the P (PO4) atoms of the two leaflets, weighted by distance to the probe. We also track the neighboring lipids for upper and lower leaflet separately and use the 10 nearest neighbors, respectively. For the trajectories, to prevent the switching of rankings of the lipids, we calculate the time-averaged distance to the probe to define the identity and rank of the 10 most important lipids on the upper and lower leaflet, respectively. Lastly, we also track the $10 \times 10$ distance matrix between 10 central atoms equivalent to the 10 beads of Martini (and the 10 beads in case of Martini) of the $3 + 3$ closest lipids and the beads of the probe.

The internal state of the lipid probe seemed of minor importance. In Martini, we only considered the distances $d(C_{3A}, C_{3B})$ and $d(C_{2A}, C_{2B})$, as well as the angles $\angle C_{3A}GL_1C_{3B}$, $\angle P_{O4}GL_1C_{1A}$ and $\angle P_{O4}GL_2C_{1B}$, which together are able to describe a split of the two lipid tails. A detailed description of these features can be found in Table S4. We proceeded to use these improved features, which are the ones discussed in Results.

While training a neural network model with a total of 663 (667 Martini) input features, we added L2 regularization to prevent over-fitting, allowing for a larger number of training epochs. Since this may inevitably punishes strong nonlinearities in the model, we then took the learned reactive flux vector $\boldsymbol{v} = \nabla\phi/|\nabla\phi|$, projected the data onto $\boldsymbol{v} \cdot \boldsymbol{x}$, and then again trained a 3-dimensional committor model using only $z$, $\xi_P$ for atomistic MD and $\theta$ for Martini MD, and $\boldsymbol{v} \cdot \boldsymbol{x}$ as inputs to test the consistency of the linear model. In Results, we discuss and contrast the resulting high and low-dimensional committor models.

### Acknowledgements
This work was supported by the Max Planck Society. We thank Balázs Fábián, Hendrik Jung, Stefan Schäfer and Jürgen Köfinger for discussion and comments on the manuscript, Vida Lotufo for assistance with analysis of DSPC lipids and the Max Planck Computing and Data Facility for computational resources.

### Author contributions
G.H. conceived the project, M.P. performed the MD simulations, M.P. analyzed the data, M.P and G.H wrote the manuscript, G.H. supervised the project.

### Competing interests
The authors declare no competing interest.

### Materials & Correspondence
Gerhard Hummer

# AI-guided transition path sampling of lipid flip-flop and membrane nanoporation
# Supplementary Information

Matthias Post* and Gerhard Hummer†

*Max Planck Institute of Biophysics, 60438 Frankfurt am Main, Germany*

(February 17, 2025)

*email: matthias.post@biophys.mpg.de
†email: gerhard.hummer@biophys.mpg.de

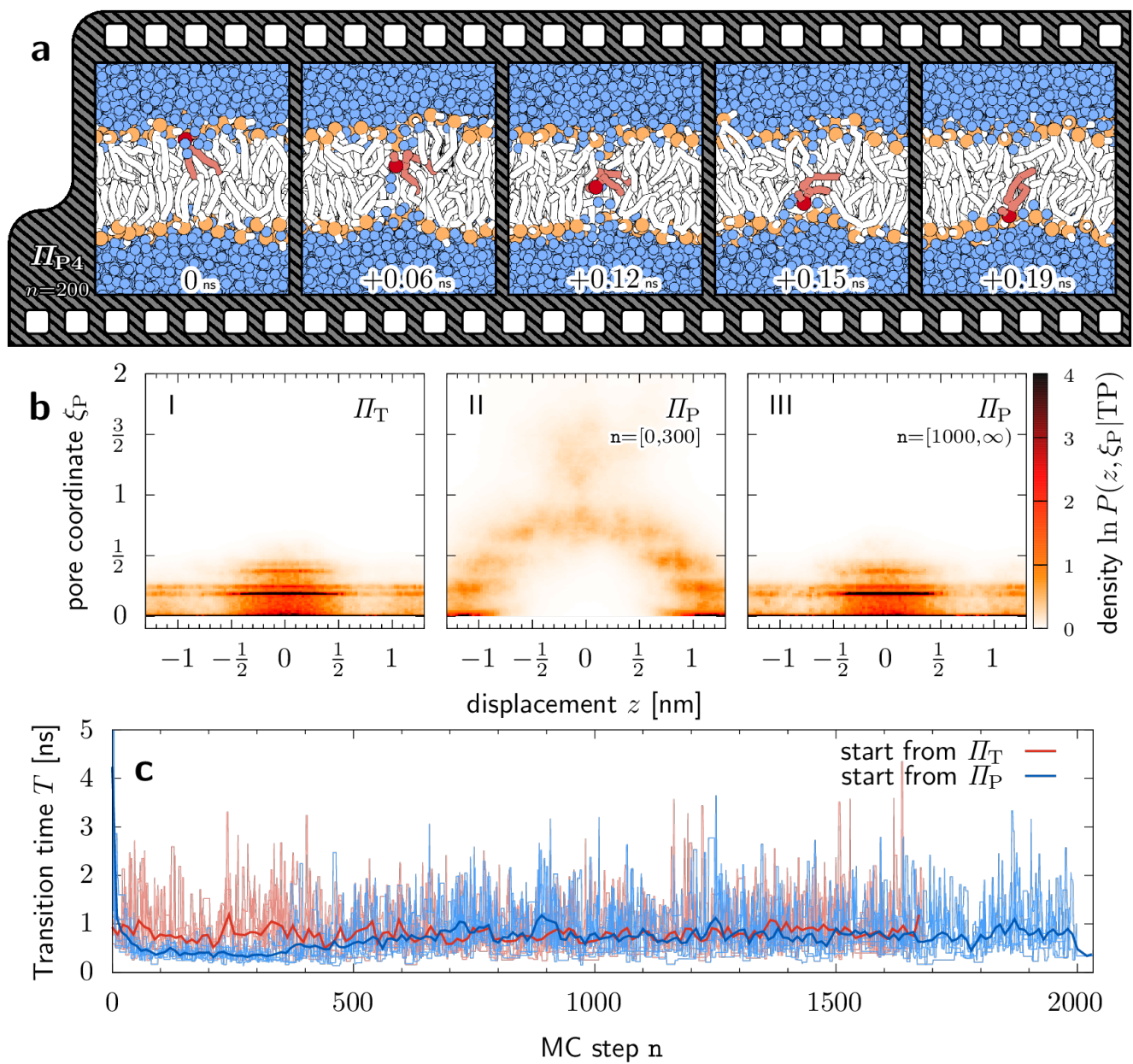

**Fig. S1. Intermediate flip-flop mechanism of Martini DMPC lipids.** (a) Exemplary trajectory during the intermediate ($n \approx 200$) interval of one of the $\Pi_P$ MC chains. Snapshots show representative intermediate conformations. Water beads in blue, lipids in white, PO4 beads orange, probe lipid in red. The box is sliced in half to see the inside. (b) $k$-neighrest neighbor TP density estimate (with $k = 500$) along transversal displacement $z$ and pore reaction coordinate $\xi_P$. We compare the TPS MC chains starting in $\Pi_T$ (bI), with those starting in $\Pi_P$, split into beginning ($n \in [0, 300]$, bII) and the end ($n \in [0, \infty)$, bIII) of the MC chain. (c) Transition times evolving with the MC chains, comparing tunnel ($\Pi_T$, red) and pore mechanism ($\Pi_P$, blue). Dark colors show an average over the faint samples, smoothed over 10 MC steps.

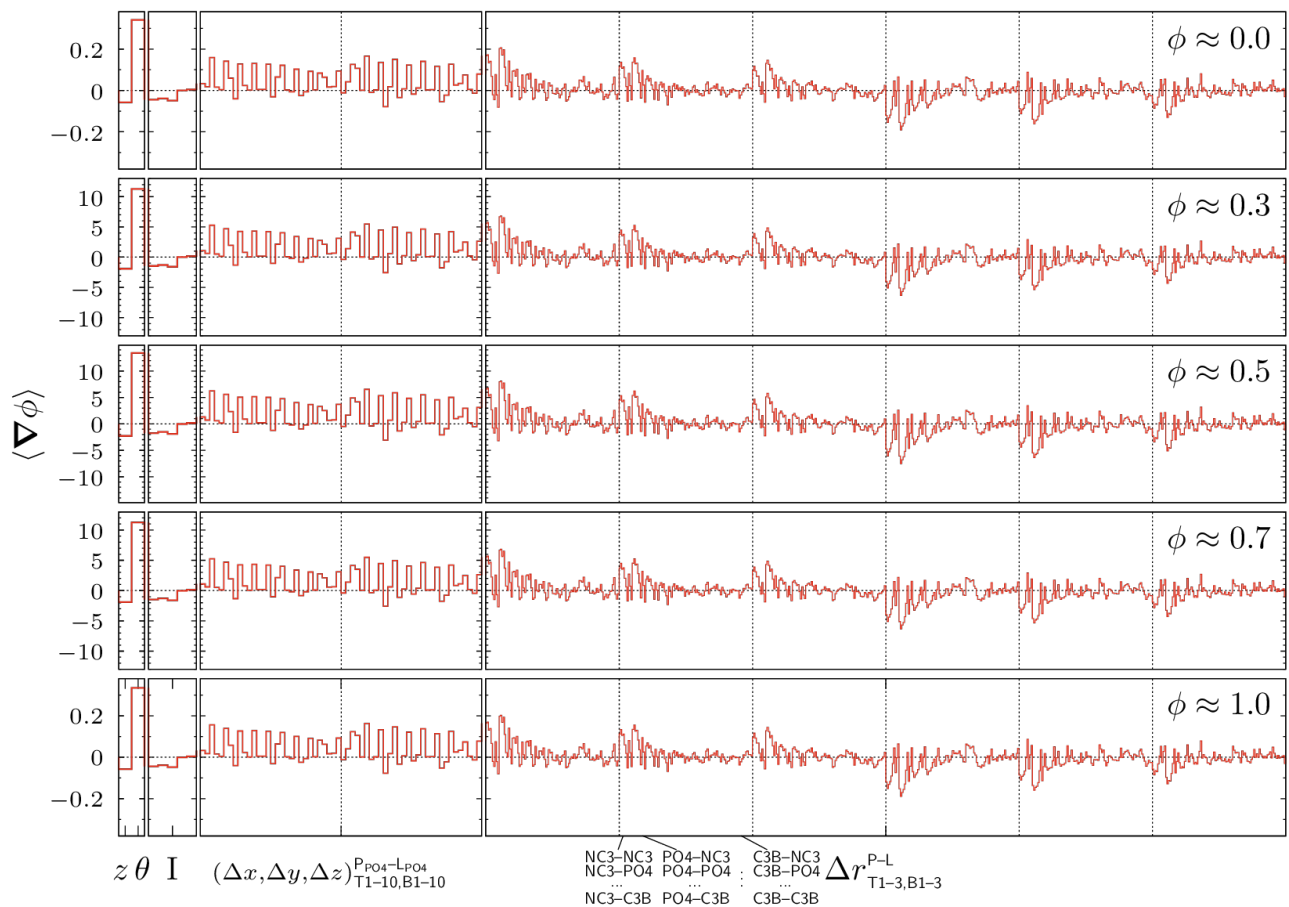

Fig. S2. **Features of committor model for flip-flop of Martini DMPC lipids**. We compare the average gradient of the committor conditioned to specific committor values. I.e., we extract all structures of the transition path ensemble in the window $\phi \pm \Delta\phi$, with $\Delta\phi = 0.01$, get $\nabla\phi$ from the network model, and then average over all frames of that window. $\nabla\phi$ was averaged over 10 models learned in each fold for cross validation. Features are grouped along the horizontal axis as indicated.

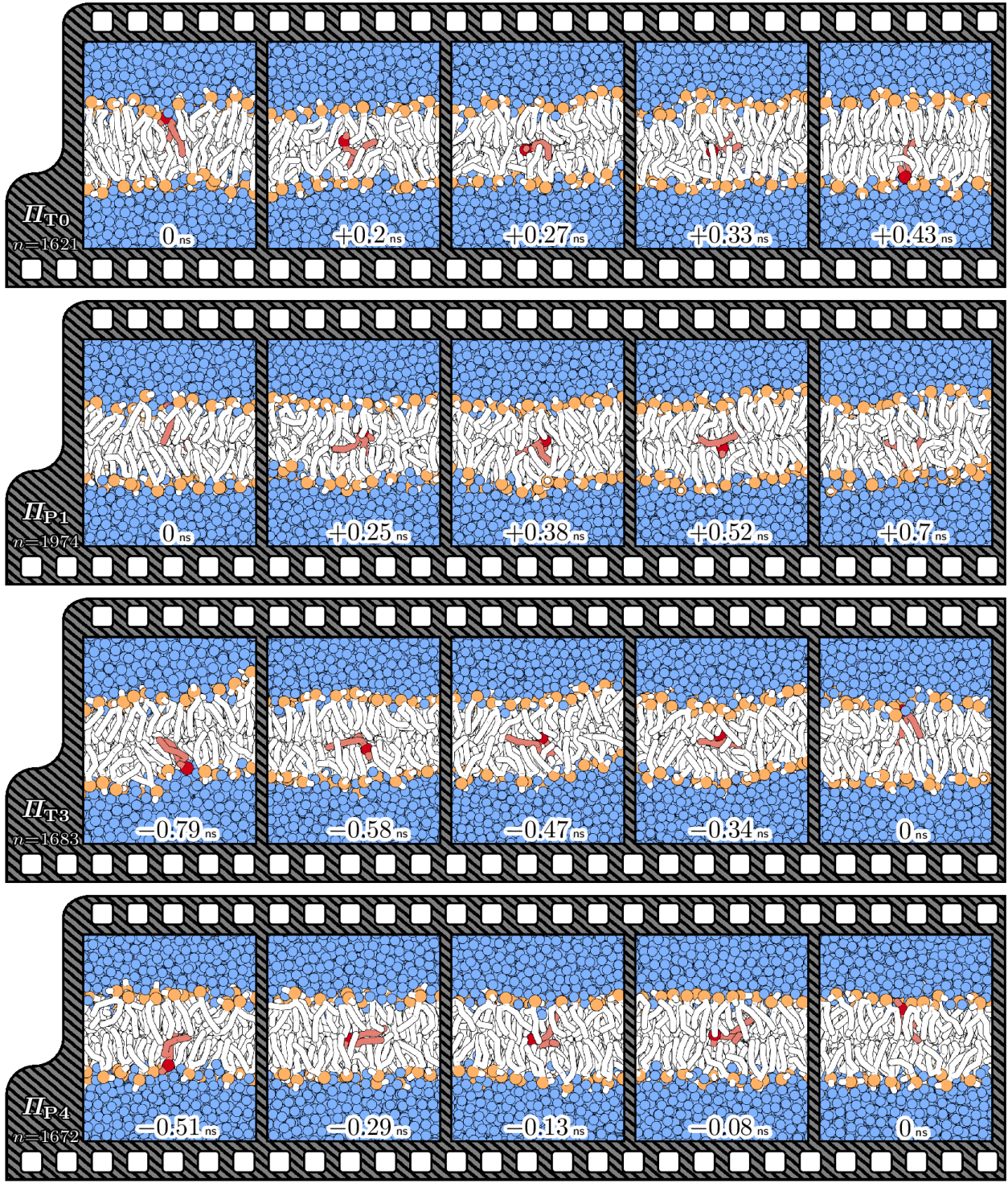

**Figure S3: Flip-flop mechanism of Martini DMPC lipids**. Snapshots of exemplary flip-flop transition trajectories of two of the samplers traversing $\mathcal{U} \to \mathcal{L}$ from a 225/225 lipid distribution to 224/226, and two in the opposite $\mathcal{L} \to \mathcal{U}$ from 224/226 to 225/225. We pick frames close to the calculated committor values $\phi = 0.0, 0.3, 0.5, 0.7$ and $1.0$, with time shown with white contour. Water and ion beads in blue. Water beads in blue, lipids in white, PO4 beads orange, probe lipid in red. The box is sliced in half to see the inside.

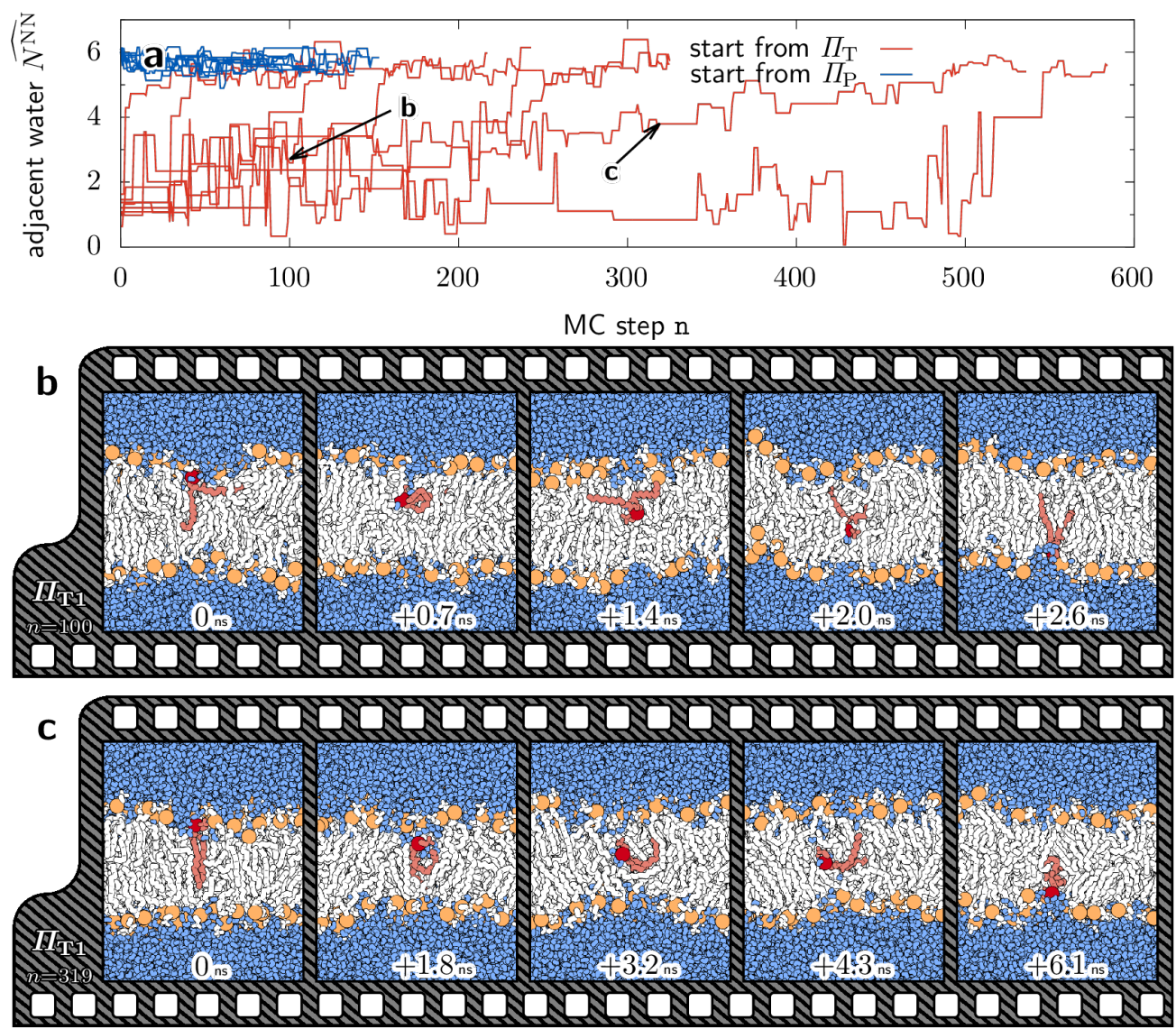

**Figure S4: Intermediate flip-flop mechanism of Charmm36 DMPC lipids.** (**a**) Average number $N^{\mathrm{NN}}$ of water beads adjacent to lipid probe, as a function of TPS MC step, comparing samplers starting in the tunnel ($\Pi_{\mathrm{T}}$, red) and in the pore mechanism ($\Pi_{\mathrm{P}}$, blue). (**b**) Exemplary trajectory of a $\Pi_{\mathrm{T}}$ sampler at the start of the chain, where $\hat{\xi}_{\mathrm{P}} < 0.3$. (**c**) Exemplary transition during of the intermediate nucleation transition from $\Pi_{\mathrm{T}}$ to $\Pi_{\mathrm{P}}$, where $0.3 < \hat{\xi}_{\mathrm{P}} < 0.9$. Snapshots show representative intermediate conformations. Water and ions in blue, lipids in white, probe lipid in red. The box is sliced in half to see the inside.

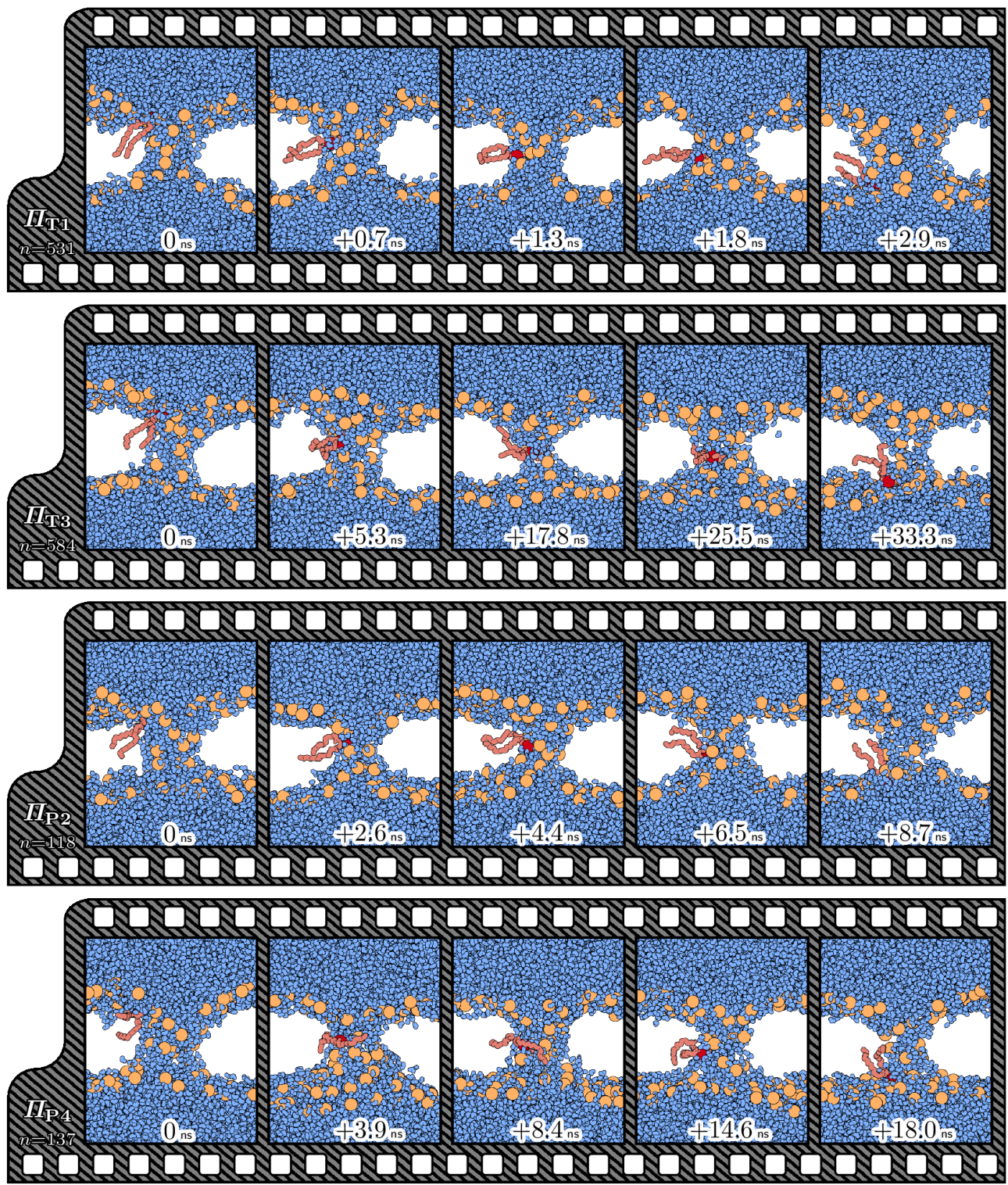

**Figure S5: Flip-flop mechanism of Charmm36 DMPC lipids.** Snapshots of exemplary transition trajectories, from the last trajectories of two of the samplers starting in $\Pi_\text{T}$ and $\Pi_\text{P}$, each. We pick frames close to the committor values $\phi = 0.0, 0.3, 0.5, 0.7$ and $1.0$, not showing excursions of the lipid probe. Water and ions in blue, lipids in white, probe lipid in red. The box is sliced in half to see the inside.

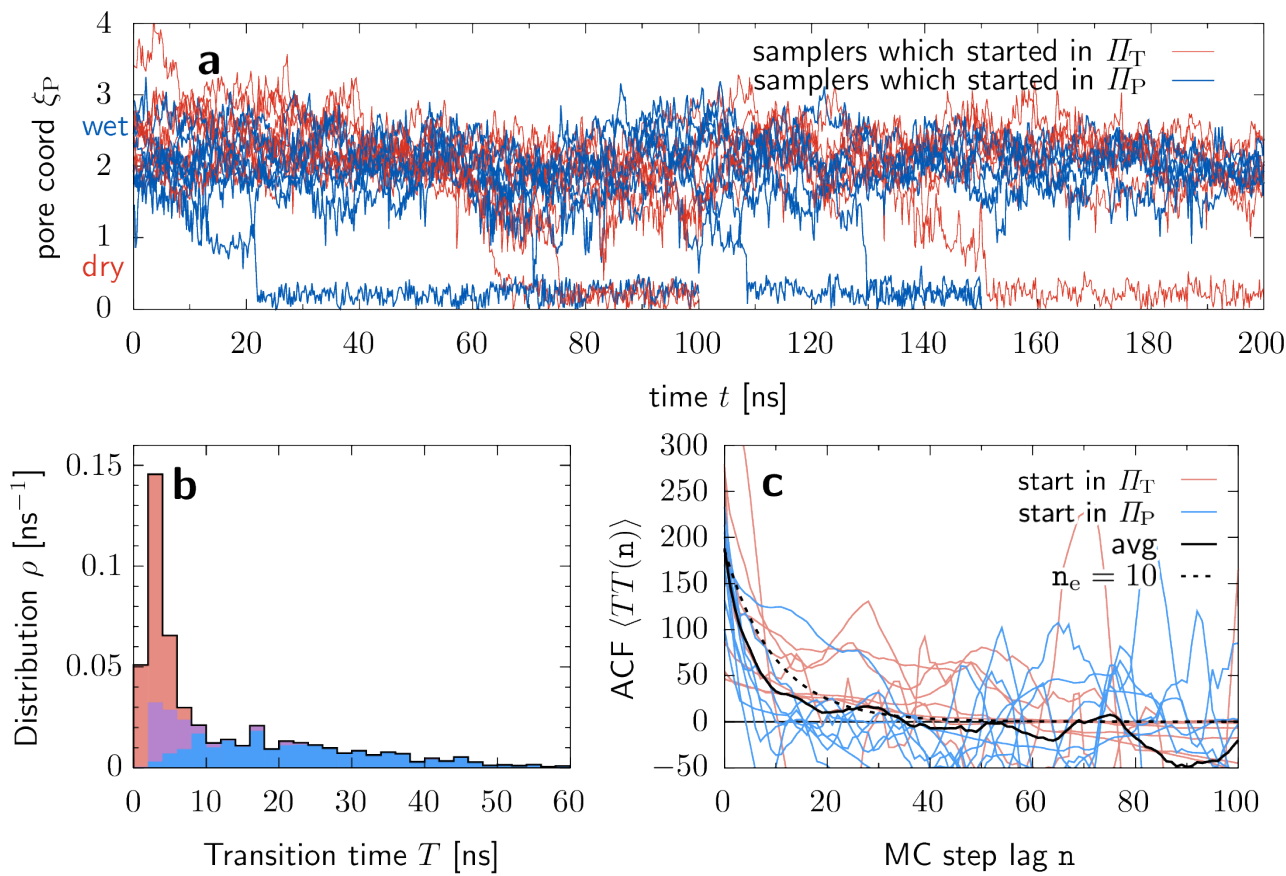

**Figure S6: Pore lifetime compared to flip-flop transition time**. (**a**) Time evolution of unbiased simulations starting from shooting point structures of the last MC step. Simulations run either a maximum of 200 ns or until collapse of the pore. We measure the rate of pore closing by counting the fraction of trajectories which did finish before the 200 ns, and divided by the aggregate simulation time to get a maximum-likelihood estimate for the closing rate of about $2.326\,\mu s^{-1}$. In the simulations with open pores, lipids were flip-flopping with a rate of $\approx 93\,\mu s^{-1}$, suggesting around 15 lipid flip-flops during the pore lifetime. (**b**) Histogram of flip-flop transition time distribution, comparing trajectories utilizing $\Pi_T$ (red), $\Pi_P$ (blue), and the intermediate nucleation transition mechanism (violet). (**c**) Auto-correlation function (ACF) of transition times between samples along the Markov chain, showing the individual sampler, as well as the average ACF (black). A bootstrap average of exponential fit parameters suggests a decorrelation "time" of $10 \pm 2$ MC steps (black dashes).

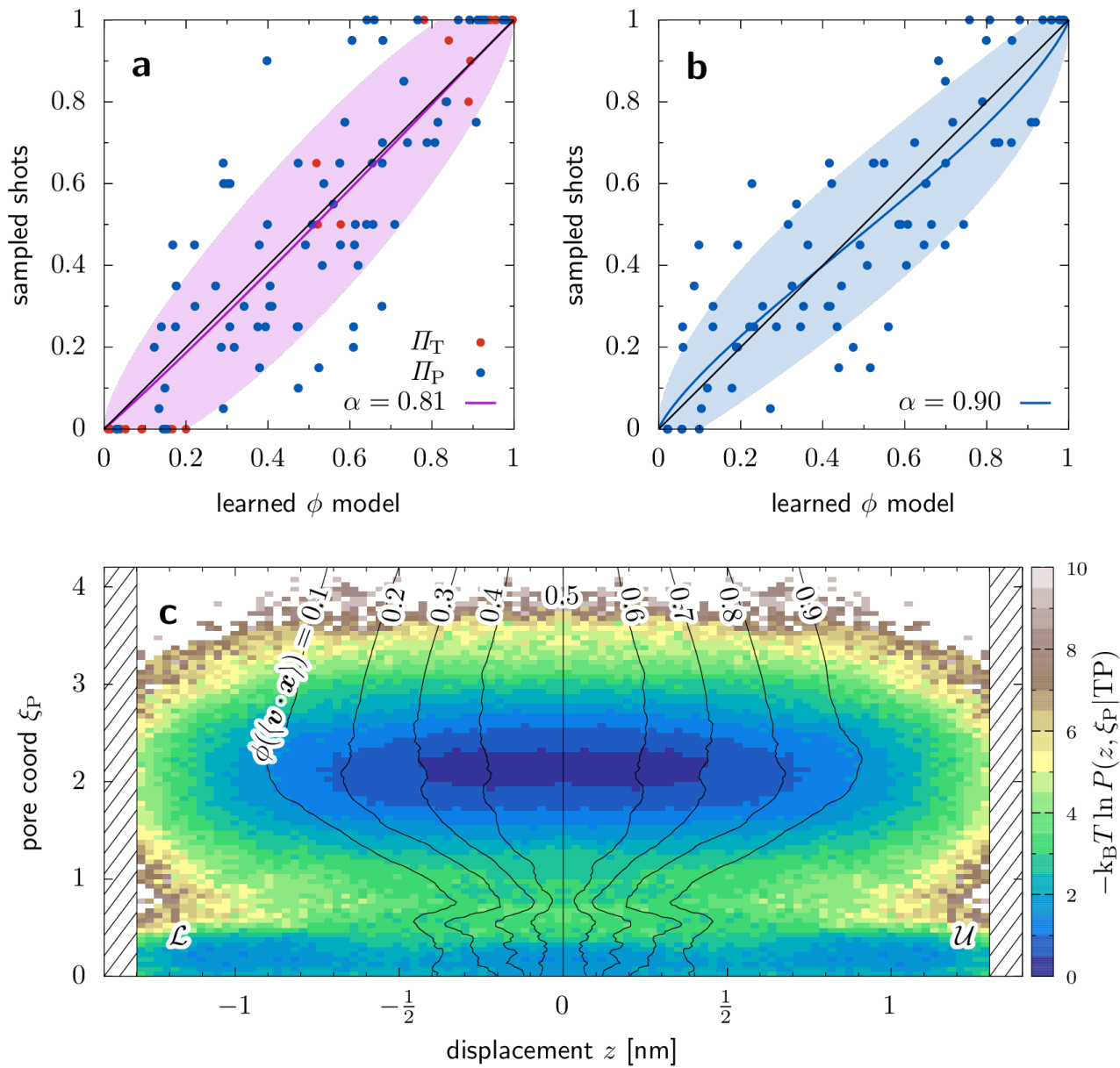

**Figure S7: Quality assessment of the Charmm36 DMPC TPE and committor estimate.** (**a,b**) Cross validation of all-atom committor model via committor shots. We picked at random 5 configurations close to $\phi_i = 0.05 \times i$ and initiated 20 trajectory shots from each. (**a**) Model trained on whole data set, showing SPs resulting in a $\Pi_T$ transition (red) and in a $\Pi_P$ transition (blue). To identify possible systematic errors in the model, we transform the learned and sampled $\phi$ to the logits $q = \ln[\phi/(1-\phi)]$, and fit a linear ansatz function (see Methods of the main text). The error band of $\pm 1\sigma$ is then drawn (shading) according to the accuracy model (see Methods). (**b**) Model trained only on the $\Pi_P$ data. (**c**) Symmetrized distribution of sampled transition along transversal displacement $z$ and pore reaction coordinate $\xi_P$. We symmetrize by the mirror image at $z = 0$, both the density, here shown as a histogram, and the iso-lines of the committor (black lines) at the average feature vector parallel to the flux, $\boldsymbol{v} \cdot \boldsymbol{x}$.

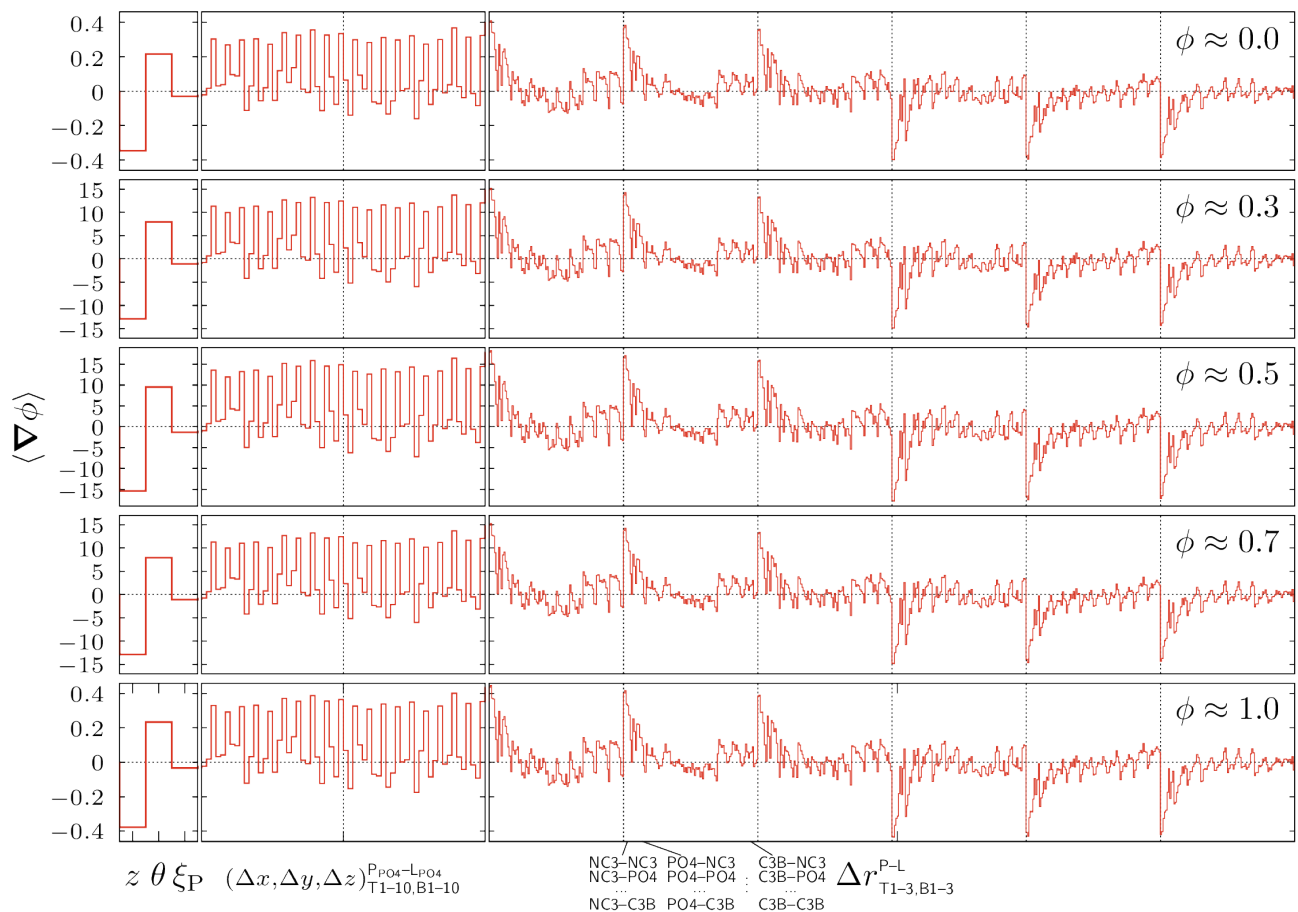

**Figure S8: Features of committor model in the all-atom DMPC lipid model**. We compare the average gradient of the committor, conditioned to specific committor values. I.e., we extract all structures of the transition path ensemble in the window $\phi \pm \Delta\phi$, with $\Delta\phi = 0.01$, get $\nabla\phi$ from the network model, and then average over all frames of that window. $\nabla\phi$ was averaged over 10 models learned in each fold for cross validation.

# TPS of Charmm36 DMPC pore nucleation

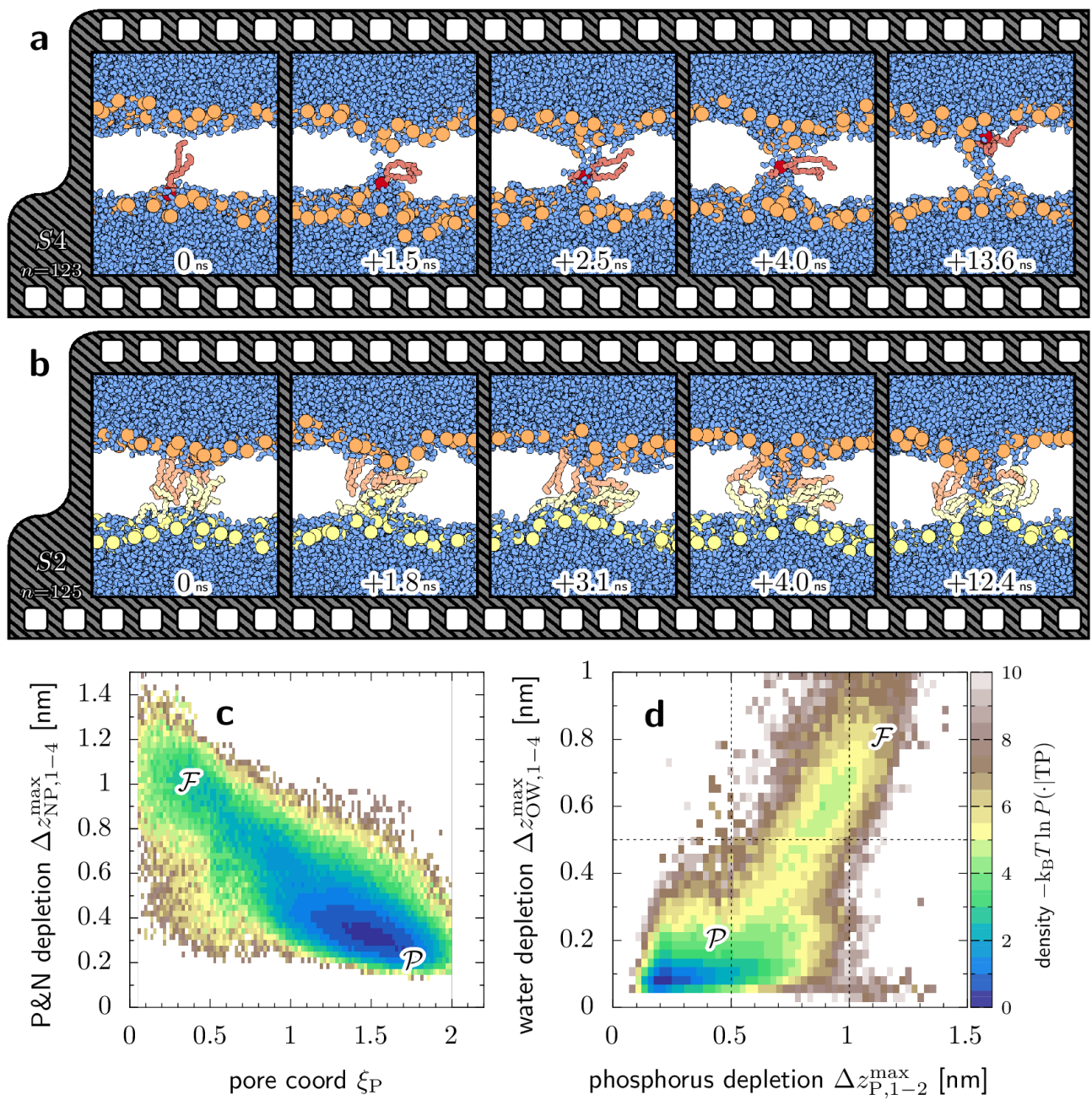

**Figure S9: Role of water and lipids during early pore nucleation.** (**a**) Displacement of the lipid closest to the midplane (red) during pore nucleation, showing one rare occasion of successful flip-flop before the pore fully expanded. DMPC phosphorus atoms are shown in orange, water ions in blue. (**b**) Another nucleation TP, where no lipid from neither the upper (orange) nor the lower (yellow) leaflet is flipping in time. (**c**) Transition path ensemble projected onto pore reaction coordinate $\xi_P$ from Refs. 1,2, compared to the largest depletion of phosphor and nitrogen, $\Delta z_{NP,1-4}^{max}$, as described in Ref. 4. (**d**) Depletion of phosphorus, $\Delta z_{P,1-2}^{max}$, compared to that of water, $\Delta z_{OW,1-4}^{max}$, resolving two stages–water first, lipids second–of pore formation.

# TPS of Charmm36 DSPC lipid flip-flop

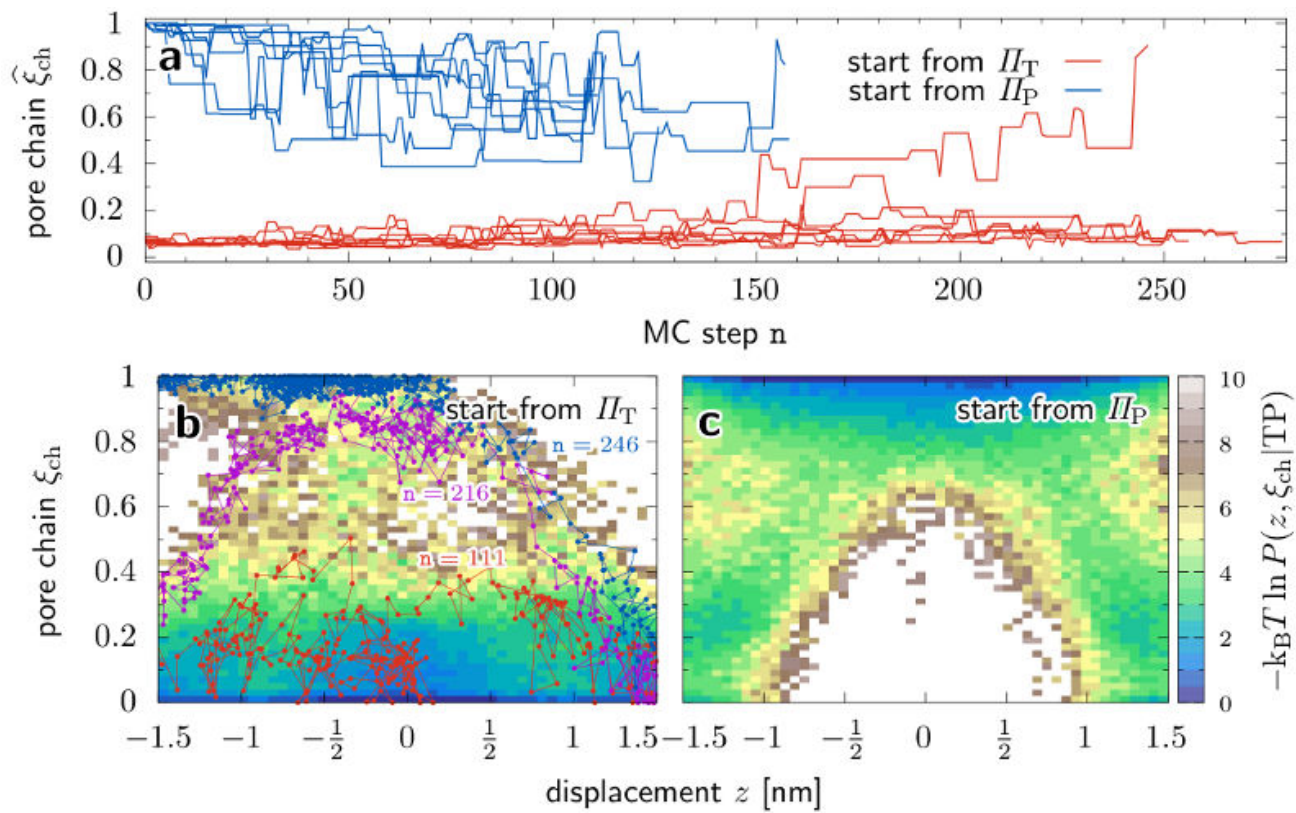

**Figure S10: DSPC lipid flip-flop.** We use the same methodology as for the Charmm36 DMPC lipids to sample the TPE of DSPC membranes. (**a**) MC chains starting from a wet ($\Pi_P$) and a dry mechanism ($\Pi_T$), showing the time-averaged pore-chain coordinate $\hat{\xi}_{ch}$ of Refs. 1, measuring the fraction of the membrane at the nucleation center already occupied by water. (**b,c**) TPE projected onto the transversal displacement $z$ and $\xi_{ch}$ in TP samplers starting from $\Pi_T$ (**b**) and $\Pi_P$ (**c**). In (**b**), we highlight the trajectories of the sampler transitioning towards the pore mechanism with dots and lines. The depletion of the low-free energy region moving from the midplane ($z = 0$ nm, $\xi_{ch} \approx 1$, blue) towards one of the leaflets ($z = \pm 1.5$ nm) highlights that frequently the pore closes as soon the probed lipid finished flipping.

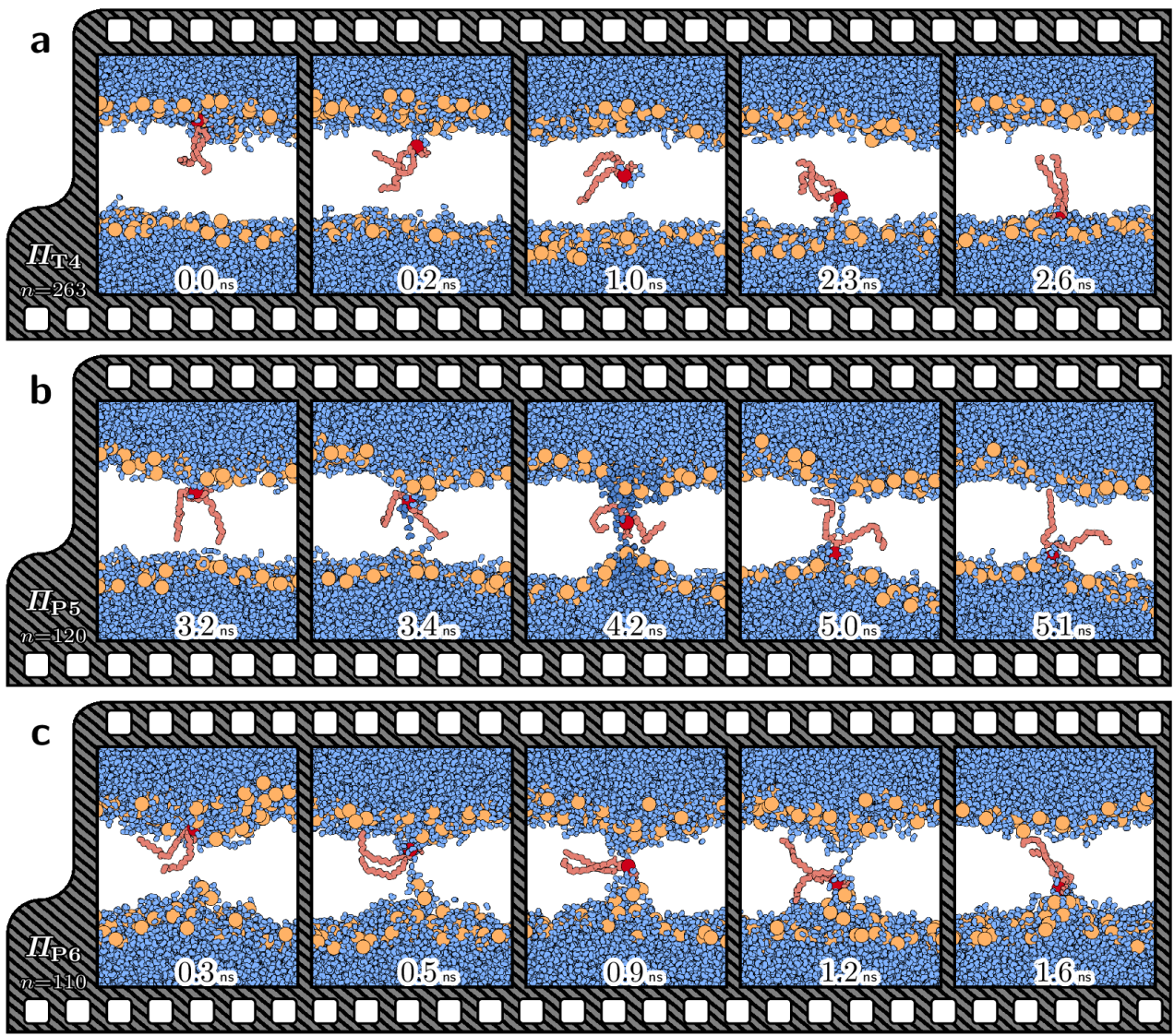

**Figure S11: Exemplary DSPC flip flop transitions**. Lipid probe shown in red, DSPC phosphorus atoms are shown in orange, water ions in blue. (**a**) Lipid probe dragging water into the bilayer surrounding its head. (**b**) The formation of a water thread before and after the flip-flop event, with the formation of a cone-like structure at 4.2 ns. We color the water molecules close to the probe's head at that time in darker blue to indicate if and how fast the water molecules traverse the bilayer. (**c**) Another example of the dominant transition mechanism of samplers in the $\Pi_P$ channel, showing how fast the pore is closing after flip-flop, as well as the initial local thinning.

# Supplementary Methods

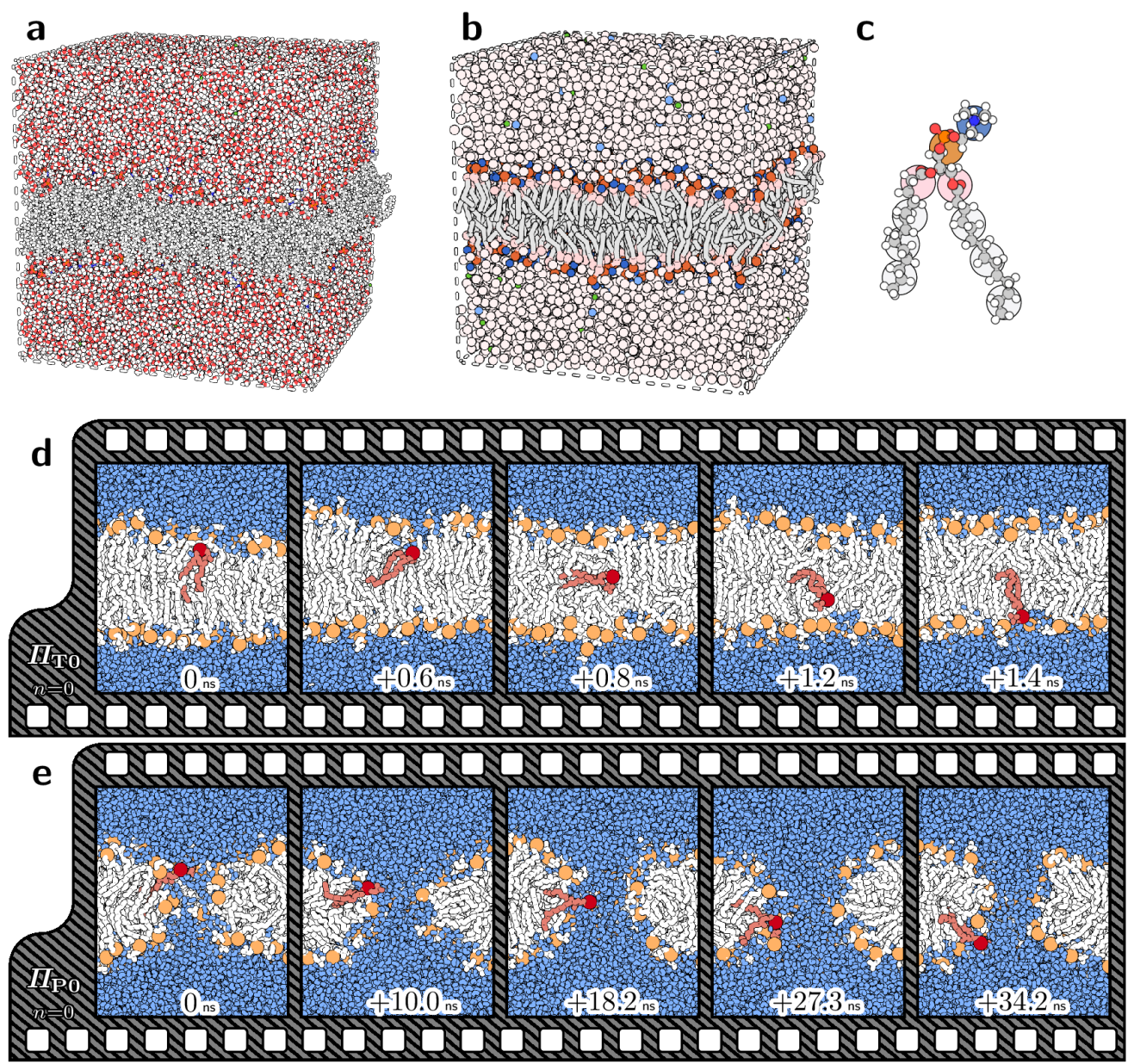

**Figure S12: Simulation setup.** (**a**) All-atom and (**b**,**c**) Martini coarse-graining. System comprising 450 DMPC lipids at 310.15 K and 1 bar. (**d**) Exemplary flip-flop transition with water kept from entering the bilayer via flat-bottomed restraint in $z$. (**e**) The same with water pore kept open via cylindrical flat-bottomed restraint of the lipids, with increasing radius along the lipid tails.

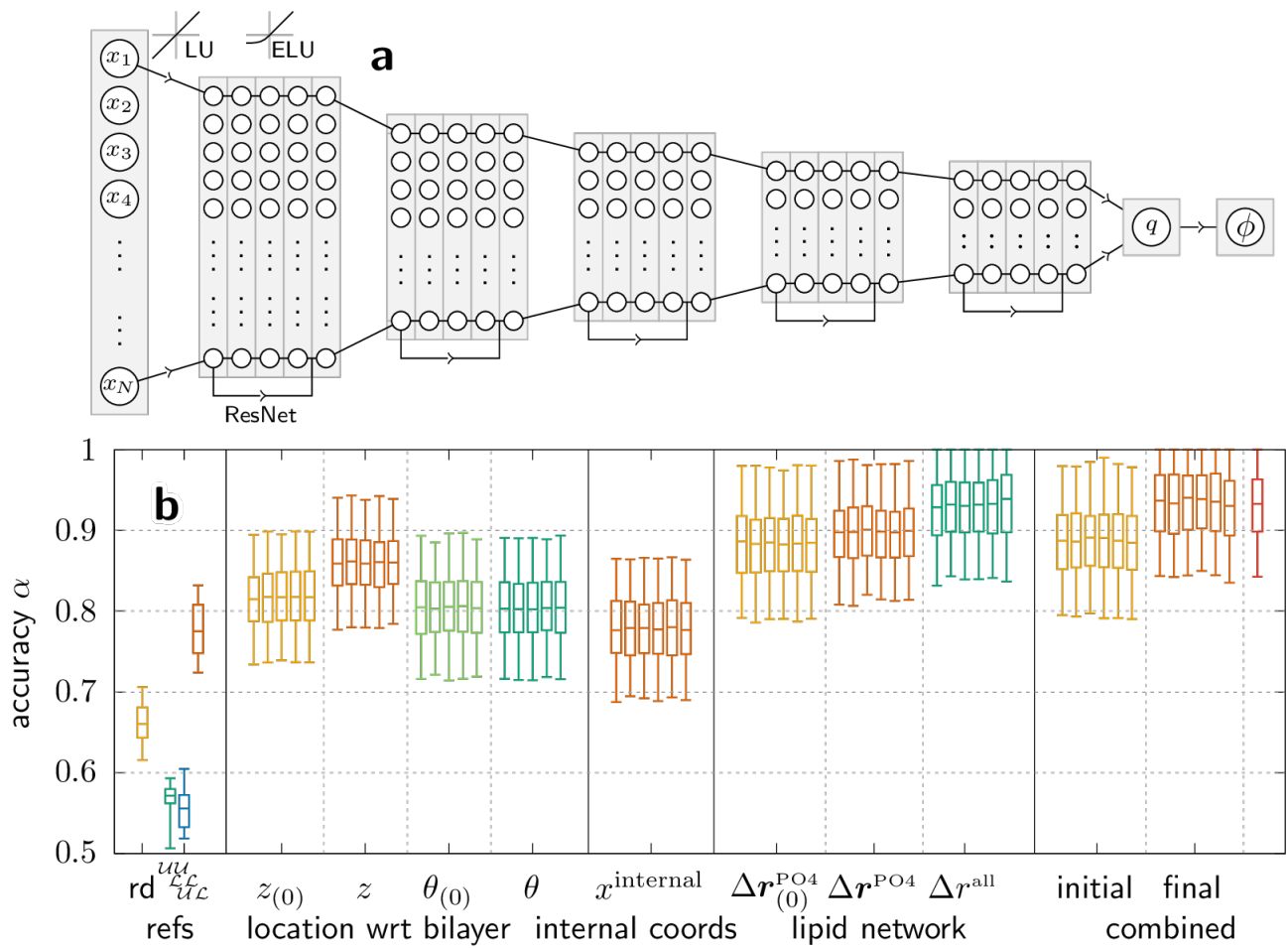

**Figure S13: Test of accuracy of various network models.** (a) General network architecture used in this work. We feed in $N$ input features $\boldsymbol{x}$, which are successively compressed via a linear unit (LU), after which a 5-layered ResNet[3] unit with exponential linear unit (ELU) activation function. The final one-dimensional output is then soft-maxed by an "expit" function. (b) Network accuracies. For reference, we guess the committor with uniform random numbers (rd) or always predict the same transition, $\phi_{\mathcal{UU}} = 0$, $\phi_{\mathcal{LL}} = 0$, or $\phi_{\mathcal{UL}} = 0.5$, with $\mathcal{U} \to$ upper and $\mathcal{L} \to$ lower leaflet. We compare the transversal displacement w.r.t. the midplane defined by all PO4 beads ($z_{(0)}$, used by the original 66d network) with one defined by the lipid tail C beads, weighted by distances to probe ($z$). The tilt angle ($\theta_{(0)}$), defined by the $z$-axis is compared with a definition weighting the PO4 beads to calculate the membrane normal ($\theta$). For the internal state we feed in a list of its distances, angles and dihedrals ($x^{\text{internal}}$). The lipid network description compares the old definition of the relative $x$, $y$ and $z$ coordinates of the next-neighboring PO4 beads ($\Delta \boldsymbol{r}^{\text{PO4}}_{(0)}$) with one sorting lipids in the upper and lower leaflet separately ($\Delta \boldsymbol{r}^{\text{PO4}}$). We also take the first three lipids from the upper and lower leaflet and compute all distance combinations between all beads with the probe ($\Delta r^{\text{all}}$). Last, we combine these improved descriptors for a final committor model (see Table S4). We estimate the accuracy, Eq. (2), by bootstrapping: With 10 folds, we split the data into 0.9 training and 0.1 validation (always leaving one of the ten samplers out), then train the network using a number of epochs, which minimizes the loss of the validation set. We then make 100 times a bootstrap sample from the validation set with repetition and calculate $\alpha$. Boxes show the median and 50% of the data, whiskers 95%. Boxes of same colors show same input features but different bottleneck architectures. The last box on the right shows the final network, which uses an additional L2 regularization term.

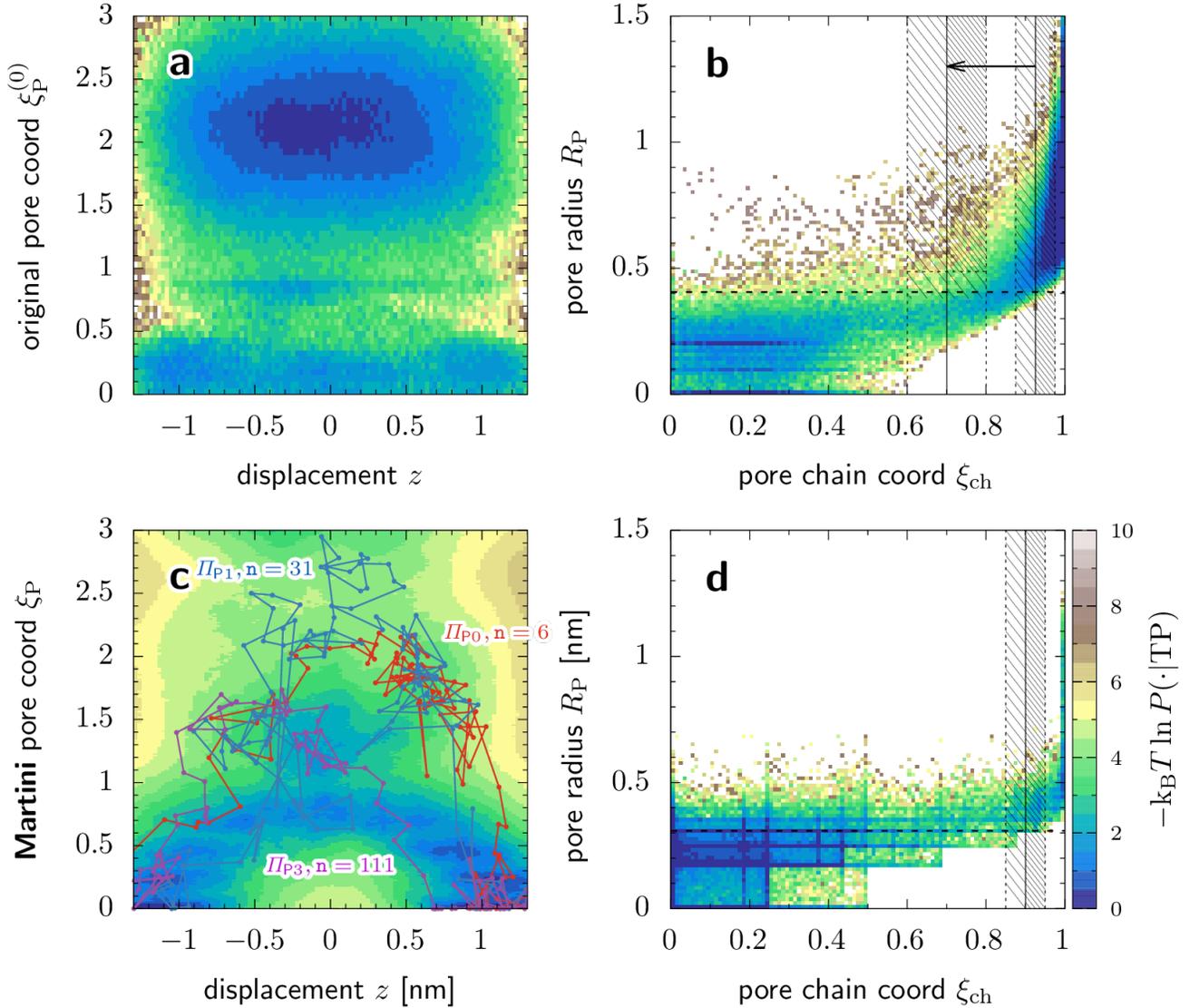

**Figure S14: Calibration of the pore reaction coordinate**. (**a,b**) TPS data of the all-atom Charmm36 DMPC lipids, projected onto coordinates describing the water pore. (**a**) The original definition in Ref. 2, $\xi_P^{(0)}$, shows an artificial meta-stable state due to its definition. (**b**) TPE projected onto the constituting coordinates, showing the old "switch" value $\xi_{ch}^s = 0.925$ up to which only the chain coordinate $\xi_{ch}$ is contributing to $\xi_P$, and the pore radius at that threshold, $R_0 = 0.405$. We instead use

$$\xi_P(r) := \xi_{ch}(r) + H_\varepsilon[\xi_{ch}(r) - \xi_{ch}^s] \times \Theta_\delta\left[\frac{R_P(r) - R_0}{R_0}\right], \quad \Theta_\delta(x) := \begin{cases} \delta e^{\frac{x-\delta}{\delta}} & \text{for } x < \delta, \\ x & \text{else.} \end{cases}$$

with $\xi_{ch}^s = 0.7$, $R_0 = 0.405$ nm, $\varepsilon = 0.1$ and $\delta = 0.2$, which removes the artifact. (**c,d**) TPS data of the Martini DMPC lipids, projected onto coordinates describing the water pore. We use the data starting from $\Pi_P$ to calibrate the parameters for $\xi_{ch}$, $R_P$ and $\xi_P$ to be able to describe the observed water defects: 4 subdivisions, $z_{mem} = 1.8$ nm, $R_{cyl} = 1.0$ nm, $D = 1.2$ nm, $\xi_{ch}^s = 0.9$, $R_0 = 0.38$ nm. (**c**) $k$-NN density of all data showing also the initial shrinking and collapse. We show exemplary trajectories at that point in the MC chain. (**d**) Data projected onto the constituting $\xi_{ch}$ and $R_P$. The sharp peaks in the density show how many water beads are in the midplane, motivating to use a small number of subdivisions. We then choose the switching region at the point where $R_P$ starts to rise above its plateau for small $\xi_{ch}$.

**Table S1.** Input features used for the committor model of lipid flip-flop trained during AIMMD.

| | | Initial descriptors |
|---|---|---|
| Category | Name | Description |
| location w.r.t. midplane | $z_{(0)}$ | Transversal displacement. We define the mid-plane as the center of all P (PO4) atoms. $z$ is the distance of the probe's P (PO4) in $z$-direction of the box w.r.t. that midplane. |
| | $\theta_{(0)}$ | Tilt angle. We use the average distance vector of each atom of the probe to its center of mass and calculate the scalar product with the $z$-axis. |
| internal coordinates | $R_G$ | Radius of gyration. Mass-weighted average of distances of each atom to the center of mass. |
| membrane | $\sigma_U$ | Upper leaflet displacement. $\sigma_U$ is calculated as the standard deviation of $z$ positions of P (PO4) atoms of the upper leaflet w.r.t. to their center. |
| | $\sigma_L$ | Lower leaflet displacement. $\sigma_L$ is calculated as the standard deviation of $z$ positions of P (PO4) atoms of the lower leaflet w.r.t. to their center. |
| lipid network | $\Delta x_{(0)}^{PO4}, \Delta y_{(0)}^{PO4}, \Delta z_{(0)}^{PO4}$ | Distance to $i$th NN lipid. On every frame of the trajectory, we sort the P (PO4) atoms by distance to the probe P (PO4). We then only keep the 20 nearest neighbors, but use all 3 signed displacements. |
| water | $N^{NN}$ | Number of water molecules next to probe. We use indicator functions of Ref. 5, $N^{NN} = \sum_i \frac{1}{2}[1 - \tanh(a(r_i - r_0))]$, where $r_i$ is the distance of the probes P (PO4) to the $i$th O (W) water atom. We use $a = 50, r_0 = 0.43$ ($a = 30, r_0 = 0.63$ Martini), corresponding to the first maximum of the RDF. |
| | $N^{NNN}$ | Number of water molecules close to probe. We use $a = 10, r_0 = 0.78$ ($a = 8, r_0 = 1.2$ Martini), corresponding to the first maximum of the RDF. |
| | $N^W$ | Number of water molecules in pore. We use the same indicator, but only with (absolute) distance in $z$-direction, with $a = 10, r_0 = 0.5$, corresponding to the width of the membrane. |

**Table S2.** Architectures of tested networks. Dropout during training shown in parentheses above arrows. Each arrow represents a linear compression, after which follows a ResNet unit, see Fig. S13a. Number of training epochs were chosen at the minimum of the validation loss. We try out all architectures for the location w.r.t. the bilayer individually. For the internal coordinates, we use all 29 together (see Table S4) For the final model, we use L2 regularization to enable a large number of epochs.

| | | List of networks -- Martini |
|---|---|---|
| Category | Name | Description |
| Reference | random | Uniform random numbers |
| | TT | Always $\phi = 0$ |
| | BB | Always $\phi = 1$ |
| | TB | Always $\phi = 0.5$ |
| location w.r.t. midplane | $z_{(0)}, z, \theta_{(0)}, \theta$ | $1 \to 10 \to 1$ network architecture |
| | | $1 \to 10 \to 10 \to 1$ |
| | | $1 \to 5 \to 10 \to 5 \to 1$ |
| | | $1 \to 5 \to 10 \to 10 \to 5 \to 1$ |
| | | $1 \to 5 \to 10 \to 10 \to 1$ |
| internal coordinates | | $29 \xrightarrow{0.24} 23 \xrightarrow{0.20} 18 \xrightarrow{0.16} 15 \xrightarrow{0.13} 12 \xrightarrow{0.10} 10 \to 1$ |
| | | $29 \to 1$ |
| | | $29 \xrightarrow{0.23} 22 \xrightarrow{0.18} 17 \xrightarrow{0.13} 13 \xrightarrow{0.10} 10 \to 1$ |
| | | $29 \xrightarrow{0.21} 20 \xrightarrow{0.15} 14 \xrightarrow{0.10} 10 \to 1$ |
| | | $29 \xrightarrow{0.18} 17 \xrightarrow{0.10} 10 \to 1$ |
| | | $29 \xrightarrow{0.12} 23 \xrightarrow{0.10} 18 \xrightarrow{0.08} 15 \xrightarrow{0.07} 12 \xrightarrow{0.05} 10 \to 1$ |
| lipid network | $\Delta r_{(0)}^{PO4}, \Delta r^{PO4}$ | $60 \xrightarrow{0.21} 41 \xrightarrow{0.15} 29 \xrightarrow{0.10} 20 \xrightarrow{0.07} 14 \xrightarrow{0.05} 10 \to 1$ |
| | | $60 \xrightarrow{0.22} 44 \xrightarrow{0.17} 33 \xrightarrow{0.11} 24 \xrightarrow{0.09} 18 \xrightarrow{0.07} 13 \xrightarrow{0.05} 10 \to 1$ |
| | | $60 \xrightarrow{0.19} 38 \xrightarrow{0.12} 24 \xrightarrow{0.08} 15 \xrightarrow{0.05} 10 \to 1$ |
| | | $60 \xrightarrow{0.16} 33 \xrightarrow{0.09} 18 \xrightarrow{0.05} 10 \to 1$ |
| | | $60 \xrightarrow{0.13} 24 \xrightarrow{0.05} 10 \to 1$ |
| | | $60 \xrightarrow{0.10} 41 \xrightarrow{0.07} 29 \xrightarrow{0.05} 20 \xrightarrow{0.03} 14 \xrightarrow{0.02} 10 \to 1$ |
| | $\Delta r^{all}$ | $600 \xrightarrow{0.13} 264 \xrightarrow{0.06} 116 \xrightarrow{0.03} 51 \xrightarrow{0.01} 22 \xrightarrow{0.005} 10 \to 1$ |
| | | $600 \xrightarrow{0.15} 303 \xrightarrow{0.08} 153 \xrightarrow{0.04} 77 \xrightarrow{0.02} 39 \xrightarrow{0.01} 19 \xrightarrow{0.005} 10 \to 1$ |
| | | $600 \xrightarrow{0.11} 215 \xrightarrow{0.04} 77 \xrightarrow{0.01} 27 \xrightarrow{0.005} 10 \to 1$ |
| | | $600 \xrightarrow{0.08} 153 \xrightarrow{0.02} 39 \xrightarrow{0.005} 10 \to 1$ |
| | | $600 \xrightarrow{0.04} 77 \xrightarrow{0.005} 10 \to 1$ |
| | | $600 \xrightarrow{0.06} 264 \xrightarrow{0.03} 116 \xrightarrow{0.02} 51 \xrightarrow{0.01} 22 \xrightarrow{0.002} 10 \to 1$ |
| initial model | $z_{(0)}, \theta_{(0)}, R_G,$ $\sigma_U,$ $\sigma_L, \Delta r_{(0)}^{PO4}, N^{NN},$ $N^{NN}, N^{W}$ | $68 \xrightarrow{0.22} 49 \xrightarrow{0.16} 35 \xrightarrow{0.12} 26 \xrightarrow{0.08} 18 \xrightarrow{0.06} 13 \xrightarrow{0.04} 10 \to 1$ |
| | | $68 \xrightarrow{0.20} 46 \xrightarrow{0.14} 31 \xrightarrow{0.09} 21 \xrightarrow{0.06} 14 \xrightarrow{0.04} 10 \to 1$ |
| | | $68 \xrightarrow{0.19} 42 \xrightarrow{0.12} 26 \xrightarrow{0.07} 16 \xrightarrow{0.04} 10 \to 1$ |
| | | $68 \xrightarrow{0.16} 35 \xrightarrow{0.08} 18 \xrightarrow{0.04} 10 \to 1$ |
| | | $68 \xrightarrow{0.12} 26 \xrightarrow{0.04} 10 \to 1$ |
| | | $68 \xrightarrow{0.11} 49 \xrightarrow{0.08} 35 \xrightarrow{0.06} 26 \xrightarrow{0.04} 18 \xrightarrow{0.03} 13 \xrightarrow{0.02} 10 \to 1$ |
| final model | $z, \theta, d_{C3A-C3B},$ $d_{C2A-C2B},$ $\theta_{C3A-GL1-C3B},$ $\theta_{PO4-GL1-C1A},$ $\theta_{PO4-GL2-C1B},$ $\Delta r^{PO4}, \Delta r^{all}$ | $667 \xrightarrow{0.13} 287 \xrightarrow{0.06} 124 \xrightarrow{0.02} 53 \xrightarrow{0.01} 23 \xrightarrow{0.004} 10 \to 1$ |
| | | $667 \xrightarrow{0.15} 331 \xrightarrow{0.07} 167 \xrightarrow{0.04} 81 \xrightarrow{0.02} 40 \xrightarrow{0.01} 20 \xrightarrow{0.004} 10 \to 1$ |
| | | $667 \xrightarrow{0.10} 233 \xrightarrow{0.04} 81 \xrightarrow{0.013} 28 \xrightarrow{0.004} 10 \to 1$ |
| | | $667 \xrightarrow{0.07} 164 \xrightarrow{0.02} 40 \xrightarrow{0.004} 10 \to 1$ |
| | | $667 \xrightarrow{0.04} 81 \xrightarrow{0.004} 10 \to 1$ |
| | | $667 \xrightarrow{0.07} 287 \xrightarrow{0.03} 124 \xrightarrow{0.02} 53 \xrightarrow{0.005} 23 \xrightarrow{0.003} 10 \to 1$ |

**Table S3.** Input features used for the committor model of pore nucleation trained during AIMMD.

| Initial descriptors | | |
|---|---|---|
| Category | Name | Description |
| Hub / Awasthi | $\xi_{ch}$ | Chain reaction coordinate as detailed in Ref. 1. |
| | $R_P$ | Pore radius as detailed in Ref. 2 |
| | $\xi_P$ | Pore reaction coordinate as detailed in Ref. 2 |
| Bubnis / Grubmueller | $\Delta r_1^X, \Delta r_2^X, \Delta r_3^X$ | Isotropic distance of closest, next, and next-to-next atom of group $X$ to the pore center. $X=$ all water oxygens, all phosphorus atoms, all phosphorus+nitrogen, all phosphorus+nitrogen+water oxygen, all lipid tail's carbons, and all carbons, as detailed in Ref. 4. The pore center we compute as detailed in Ref. 1. |
| | $\Delta \rho_1^X, \Delta \rho_2^X, \Delta \rho_3^X$ | Lateral distance of the same. |

|  | $\Delta z_1^X$, $\Delta z_2^X$, $\Delta z_3^X$ | Axial distance of the closest atoms. |
|---|---|---|
|  | $\Delta r_{1-2}^X$, $\Delta r_{1-3}^X$, $\Delta r_{1-4}^X$, $\Delta r_{1-5}^X$, $\Delta r_{1-10}^X$ | Isotropic distance averaged over the closest atoms as indicated, i.e., over the 2, 3, 4, 5 and 10 atoms closest to the nucleation center. |
|  | $\Delta \rho_{1-2}^X$, $\Delta \rho_{1-3}^X$, $\Delta \rho_{1-4}^X$, $\Delta \rho_{1-5}^X$, $\Delta \rho_{1-10}^X$ | Lateral distance of the same. |
|  | $\Delta z_{X1-2}^{max}$, $\Delta z_{X1-3}^{max}$, $\Delta z_{X1-4}^{max}$, $\Delta z_{X1-5}^{max}$, $\Delta z_{X1-10}^{max}$ | The closest 100 atoms to the center are sorted by their z-position. Then, we loop over pairs of 2, 3, 4, 5 and 10 atoms, respectively, and calculate the average distance to their average z-position. The "depletion" of atoms is then the respective maximum. |

**Table S4.** Input features used for the committor models trained post-simulation.

| Category | Name | Description |
|---|---|---|
| | Additional descriptors – Martini Name Description | |
| position w.r.t. midplane | $z$ | We define the mid-plane via all lipid tail carbon beads. We weigh them by their distance to the probe: We use the sigmoid of Ref. 6, $w_i = \left[1 + (2^{a/b} - 1)(\Delta r/s)^a\right]^{-b/a}$, with $a = 8$, $b = 3$, $\sigma = 2$ nm, to weigh lipids with distance $\Delta r < \sigma$ more. |
| | $\theta$ | Instead of the z-axis, we use the distance vector spanned by the center of PO4 beads of the upper leaflet and the lower leaflet. We weight these atoms again by distance to the probes P atom, as above. |
| internal coordinates | $d_{GL1-GL2}$, $d_{C1A-C1B}$, $d_{C2A-C2B}$, $d_{C3A-C3B}$, $d_{NC3-C1A}$, $d_{NC3-C2A}$, $d_{NC3-C3A}$, $d_{NC3-C1B}$, $d_{NC3-C2B}$, $d_{NC3-C3B}$ | Distance between beads of lipid probe. |
| | $\theta_{NC3-PO4-GL1}$, $\theta_{NC3-PO4-GL2}$, $\theta_{GL1-PO4-GL2}$, $\theta_{C1A-PO4-C1B}$, $\theta_{C2A-PO4-C2B}$, $\theta_{C3A-PO4-C3B}$, $\theta_{PO4-GL1-C1A}$, $\theta_{GL1-C1A-C2A}$, $\theta_{C1A-C2A-C3A}$, $\theta_{PO4-GL2-C1B}$, $\theta_{GL2-C1B-C2B}$, $\theta_{C1B-C2B-C3B}$ | Angles between beads of lipid probe. |
| | $\varphi_{NC3-PO4-GL1-GL2}$, $\varphi_{NC3-PO4-GL1-C1A}$, $\varphi_{PO4-GL1-C1A-C2A}$, $\varphi_{GL1-C1A-C2A-C3A}$, $\varphi_{NC3-PO4-GL2-C1B}$, $\varphi_{PO4-GL2-C1B-C2B}$, $\varphi_{GL2-C1B-C2B-C3B}$ | Dihedral between beads of lipid probe. |
| lipid network | $\Delta x^{PO4}$, $\Delta y^{PO4}$, $\Delta z^{PO4}$ | Lipids PO4 displacement sorted by a fixed reference. We calculate their distance to the probes PO4 over time to measure their importance weight $w_i(\Delta r)$ with $a = 20$, $b = 8$ and $\sigma = 0.7$. We then take the time average to sort by highest importance, and save the 10 most important lipids of the lower and upper leaflet, respectively. |
| | $\Delta r^{all}$ | We use the above ranking to calculate the distance matrix between all 10 beads of the probe and the $i$th ranked lipid. We save the three most important (see above) lipids of the lower and upper leaflet, respectively. |